\documentclass[aps,pra,showpacs,twocolumn]{revtex4}
\usepackage{amsmath, amssymb}
\usepackage{graphicx}
\usepackage{subfigure}

\newcommand{\re}{\mathrm{Re}}
\newcommand{\im}{\mathrm{Im}}

\begin{document}

\title{Generalized minimum-uncertainty squeezed states}

\author{E. Shchukin}
\email{evgeny.shchukin@gmail.com}
\author{W. Vogel}
\email{werner.vogel@uni-rostock.de}
\author{Th. Kiesel}
\email{thomas.kiesel@uni-rostock.de}
\affiliation{Arbeitsgruppe Quantenoptik, Institut f\"{u}r  Physik,
Universit\"at Rostock, D-18051 Rostock, Germany}

\begin{abstract}
Minimum-uncertainty squeezed states, related to a broad class of observables, are analyzed. Methods for characterizing such states are developed, which are based on numerical solutions of ordinary differential equations. As typical examples we deal with nonlinear generalizations of quadrature squeezed states and deformed nonlinear squeezed states. In this manner one may derive those squeezed states which are directly related to given observables. This can be useful for optimized measurements at a reduced level of quantum-noise.
\end{abstract}


% \submitto{\jpb}
% 
\maketitle

\section{Introduction}

It is a well-known fact that the ground-state and the vacuum noise level of a harmonic oscillator and a mode of the radiation field, respectively, is required to fulfill the Heisenberg uncertainty relation. In fact, this noise level just defines the minimum uncertainty level of the quantum noise which is required to obey the uncertainty principle. It is important to note that the uncertainty principle sets a limit for the product of the variances of two observables, such as position and momentum of a harmonic oscillator and two orthogonal field quadratures for a radiation mode. Therefore it is a natural conclusion that one may find quantum states, for which the noise in one of the chosen pair of observables is reduced below the vacuum (or ground-state) noise level, at the expense of increased noise in the other observable \cite{prd-4-1925, prd-1-3217}. Nowadays such states are usually called squeezed states, in the early days of their study also the notion two-photon coherent states was used~\cite{pra-13-2226, pla-68-321}, since the structure of the unitary operator which leads to such state is formally a two-photon generalization of the coherent displacement operator.
  
More than twenty years ago that the first successful experimentell realization of squeezed radiation fields have been published \cite{prl-55-2409}. There has been some interest in possible applications of the noise reduction in a given observable, for example in the context of interferometric detection of gravitational waves \cite{prd-23-1693, prl-47-815}. It could be shown that the squeezing effect indeed improves  
interferometric \cite{prl-59-278, prl-59-2153} and spectroscopic measurements \cite{prl-68-3020}. Very recently squeezing could be realized with a reduction of the noise power by 10 dB \cite{schnabel},
which makes the squeezed states indeed useful for gravitational wave astronomy.

Based on this encouraging progress in the generation and application of the so-called quadrature squeezed states, it is of some interest to rise the question of whether one may consider useful generalizations of  the concept of squeezed states. For example, generalization of squeezing has been proposed, which is based on the uncertainty relation of two general non-commuting observables~\cite{prl-47-709}.  
The further investigation of squeezing in such a general sense is closely connected with the minimization of the uncertainty relation of two Hermitian operators. The problem of finding such generalized squeezed  states requires to solve the minimization problem of the uncertainty relation for the 
two Hermitian operators defining the squeezing under consideration. The resulting minimum uncertainty states are then the generalized squeezed states. Generalized squeezing in this sense has been considered for some special choices of the basic operators, such as amplitude-squared squeezing \cite{pra-36-3796} and its higher-order generalizations \cite{pla-150-27, jp-36-183}, for a review see also \cite{job-4-R1}. Other generalizations of quadrature squeezing were based on the consideration of higher-order moments~\cite{prl-54-323}, and on higher powers of the annihilation/creation operators in the squeeze operator~\cite{pra-35-1659}. 

In the present paper we will consider minimum-uncertainty squeezed states for two general noncommuting observables. Pure quantum states which fulfill this requirement will be constructed as the solutions of an eigenvalue problem, which can be analytically solved only in a few special cases, including quadrature squeezing and amplitude squared squeezing.  For more general choices of the two Hermitian operators, we develop a systematic approach to find such states numerically in the Fock-Bargmann representation. This method is applied to other types of squeezed states, such as  generalized quadrature squeezed states and deformed nonlinear squeezed states. It allows one to obtain and characterize the optimized squeezed states for a chosen observable. This may provide a powerful tool for optimizing a given measurement principle with respect to the relevant level of quantum noise. 

The paper is organized as follows. In Sec.~2 we provide some useful
methodical details of the Fock-Bargman representation. The definition of the
generalized squeezed states under study is introduced in Sec.~3. 
The generalization of quadrature squeezing is studied in Sec.~4, where in the
quadrature operator the annihilation operator is replaced by a function of the
latter.
In Sec.~5 we reconsider the known effects of quadrature squeezing and
amplitude-squared squeezing from the viewpoint of our general method, in these
cases  we are able to provide analytical solutions for the corresponding squeezed states.
Deformed nonlinear squeezed states are introduced in Sec.~6, and some of their properties are analyzed. A summary and some conclusions are given in Sec.~7.

%\section{Generalized minimum-uncertainty squeezed states}

\section{Fock-Bargmann representation}

In this section we discuss the Fock-Bargmann representation of pure states and give some usefull expressions for the quantities we need below.
Any pure quantum state $|\psi\rangle = \sum^{+\infty}_{n=0} c_n |n\rangle$ can be written as the action of the operator $\psi(\hat{a}^\dagger)$, which is a function of the creation operator only, on the vacuum state
\begin{equation}\label{eq:FBr}
    |\psi\rangle = \psi(\hat{a}^\dagger) |0\rangle,
\end{equation}
where the function $\psi(z)$ is defined via
\begin{equation}
    \psi(z) = \sum^{+\infty}_{n=0} c_n \frac{z^n}{\sqrt{n!}} = \langle z|\psi\rangle e^{|z|^2/2}.
\end{equation}
This series converges producing an entire analytical function. The representation of quantum states by means of entire analytical functions according to Eq.~(\ref{eq:FBr}) is referred to as Fock-Bargmann representation \cite{fock-bargmann, bargmann-fock}. The scalar product $\langle\psi_1|\psi_2\rangle$ in this representation reads as
\begin{equation}\label{eq:FBs}
\begin{split}
    \langle\psi_1|\psi_2\rangle &= \int \psi^*_1(z) \psi_2(z) e^{-|z|^2}\,d^2z \\
    &= \sum^{+\infty}_{k=0} \frac{\psi^{*(k)}_1(0) \psi^{(k)}_2(0)}{k!},
\end{split}
\end{equation}
in particular, the normalization of a state $|\psi\rangle$ is given by the following condition:
\begin{equation}\label{eq:FBn}
    \langle\psi|\psi\rangle = \int |\psi(z)|^2 e^{-|z|^2}\,d^2z = \sum^{+\infty}_{n=0} \frac{|\psi^{(n)}(0)|^2}{n!} = 1.
\end{equation}
If an entire analytical function $\psi(z)$ satisfies a weaker condition
\begin{equation}\label{eq:FBn2}
    \mathcal{N}^{-2} = \int |\psi(z)|^2 e^{-|z|^2}\,d^2z < +\infty.
\end{equation}
then the state defined by Eq.~(\ref{eq:FBr}) is not normalized, and its normalization is the number $\mathcal{N}$ defined by Eq.~(\ref{eq:FBn2}).

A function $g(\hat{a}^\dagger)$ of the creation operator $\hat{a}^\dagger$ is the multiplication by $g(z)$ in the Fock-Bargmann representation. Due to the equality
\begin{equation}
    \hat{a} \psi(\hat{a}^\dagger) = \psi(\hat{a}^\dagger) \hat{a} + \psi^\prime(\hat{a}^\dagger),
\end{equation}
the annihilation operator $\hat{a}$ corresponds to the derivative $d/dz$. The relation (\ref{eq:no}) from \ref{app:NOa} shows that 
\begin{equation}\label{eq:FBe}
    e^{\alpha \hat{a}} \psi(\hat{a}^\dagger) |0\rangle = \sum^{+\infty}_{k=0} \frac{\alpha^k}{k!} \psi^{(k)}(\hat{a}^\dagger) |0\rangle = \psi(\hat{a}^\dagger + \alpha) |0\rangle,
\end{equation}
thereby the operator $e^{\alpha \hat{a}}$ is the shift of the argument in the Fock-Bargmann representation. The relation
\begin{equation}
    \mu^{\hat{n}} \psi(\hat{a}^\dagger) |0\rangle = \psi(\mu \hat{a}^\dagger) |0\rangle
\end{equation}
shows that the operator $\mu^{\hat{n}}$ corresponds to the scaling of the argument. The moments $\langle \hat{a}^{\dagger n} \hat{a}^m \rangle$ in the Fock-Bargmann representation are readily calculated as follows:
\begin{equation}\label{eq:FBm}
\begin{split}
    \langle \hat{a}^{\dagger n} \hat{a}^m \rangle &= \int \psi^{*(n)}(z) \psi^{(m)}(z) e^{-|z|^2}\,d^2z \\
    &= \sum^{+\infty}_{k=0} \frac{\psi^{*(n+k)}(0) \psi^{(m+k)}(0)}{k!}.
\end{split}
\end{equation}
Usually it is much easier to use the discrete versions of Eqs.~(\ref{eq:FBs}), (\ref{eq:FBn}) and (\ref{eq:FBm}) than to calculate the corresponding double integrals. Using Eq.~(\ref{eq:FBe}) one can get the following expression for the normally-ordered characteristic function $\Phi(\beta, \beta^*) = \langle e^{\beta\hat{a}^\dagger} e^{-\beta^*\hat{a}} \rangle$ of the state (\ref{eq:FBr}):
\begin{equation}
\begin{split}
    \Phi(\beta, \beta^*) &= \int \psi^*(z+\beta) \psi(z-\beta^*) e^{-|z|^2}\,d^2z \\
    &= \sum^{+\infty}_{k=0} \frac{\psi^{* (k)}(\beta) \psi^{(k)}(-\beta^*)}{k!}.
\end{split}
\end{equation}
Note that this expression is in agreement with Eq.~(\ref{eq:FBm}) since the moments $\langle \hat{a}^{\dagger n} \hat{a}^m \rangle$ can be expressed in terms of the characteristic function $\Phi(\beta, \beta^*)$ as follows:
\begin{equation}\label{eq:FBmc}
    \langle \hat{a}^{\dagger n} \hat{a}^m \rangle = \left.\frac{\partial^{n+m} \Phi(\beta, \beta^*)}{\partial\beta^n \partial(-\beta^*)^m}\right|_{\beta=\beta^*=0}.
\end{equation}
Below we need to calculate the moments of the squeezed variant of the state (\ref{eq:FBr})
\begin{equation}
    |\psi_{\mathrm{sq}}\rangle = S(\xi)|\psi\rangle = S(\xi)\psi(\hat{a}^\dagger) |0\rangle.
\end{equation}
where $S(\xi) = e^{(\xi^*\hat{a}^2-\xi\hat{a}^{\dagger 2})/2}$ is the squeezing operator. The moments of this state can be obtained from Eq.~(\ref{eq:FBmc}) and the following general relation for the characteristic function $\Phi_{\mathrm{sq}}(\beta, \beta^*)$ of the squeezed variant of any quantum state:
\begin{equation}\label{eq:FBsc}
    \Phi_{\mathrm{sq}}(\beta, \beta^*) = \Phi(\beta^\prime, \beta^{\prime*}) e^{-|\nu|^2 |\beta|^2 - \mu \re(\nu^*\beta^2)},
\end{equation}
where $\beta^\prime = \mu \beta+\nu \beta^*$. In some examples considered below there is no closed analytical expression for the characteristic function (to our knowledge), but in all examples it is possible to get such expressions for the moments using Eq.~(\ref{eq:FBm}) (though sometimes these expressions are really huge).

\section{Generalized squeezing}

Let us now consider the definition of generalized squeezed states in more detail. The calculation of the properties of such states will be reduced to the problem of solving ordinary differential equations. 
Some examples will be studied in the next section.

The general uncertainty relation for two Hermitian operators $\hat{F}$ and $\hat{G}$ of equal dimension reads as follows:
\begin{equation}\label{eq:GSur}
    \Delta F \Delta G \geq \frac{1}{2} |\langle[\hat{F}, \hat{G}]\rangle|,
\end{equation}
where $\Delta A = \langle(\Delta \hat{A})^2\rangle^{1/2}$ is the dispersion of $\hat{A}$. In this work we study the states which minimize this uncertainty relation, i.e. the states which satisfy the equality
\begin{equation}\label{eq:GSu}
    \Delta F \Delta G = \frac{1}{2} |\langle[\hat{F}, \hat{G}]\rangle|.
\end{equation}
Unless $\Delta F = \Delta G$, exactly one of the following inequalities  is valid:
\begin{equation}\label{eq:GSc}
    (\Delta F)^2 < \frac{1}{2}|\langle[\hat{F}, \hat{G}]\rangle| \quad \mathrm{or} \quad
    (\Delta G)^2 < \frac{1}{2}|\langle[\hat{F}, \hat{G}]\rangle|.
\end{equation}
A state that satisfies any of these inequalities was called generalized squeezed state (\cite{prl-47-709, oc-61-432}).

In this work we deal with pure states only. Any solution $|\psi\rangle$ of the eigenvalue problem 
\begin{equation}\label{eq:GSevp}
    (\hat{F}+i\lambda\hat{G})|\psi\rangle = \beta |\psi\rangle,
\end{equation}
where $\lambda$ is a positive real number and $\beta$ is arbitrary complex, also satisfies Eq.~(\ref{eq:GSu}) cf. \cite{schiff}. The relation between the dispersions $\Delta F$ and $\Delta G$ for such a state reads as
\begin{equation}
    \Delta F = \lambda \Delta G,
\end{equation}
so the parameter $\lambda$ plays the role of the degree of squeezing. For $0 < \lambda < 1$ the first of the inequalities (\ref{eq:GSc}) is satisfied, and for $\lambda > 1$ the second one is. The solutions of Eq.~(\ref{eq:GSevp}) for $\lambda=1$ are unsqueezed (in the generalized sense under consideration). Nevertheless the resulting state can be a nonclassical one.

In general, it is impossible to solve Eq.~(\ref{eq:GSevp}) analytically, but one can rewrite it as an ordinary differential equation, whereby making it possible to solve it numerically. To transform Eq.~(\ref{eq:GSevp}) to an ordinary differential equation note that for a coherent state $|\alpha\rangle$ we have the relations
\begin{equation}\label{eq:GSa}
    \langle\alpha|\hat{a}^\dagger = \alpha^*\langle\alpha|, \quad
    \langle\alpha|\hat{a} = \left(\frac{\alpha}{2}+\frac{\partial}{\partial\alpha^*}\right)\langle\alpha|.
\end{equation}
We assume that the operators $\hat{F}$ and $\hat{G}$ are written in the normally-ordered form. Using the relations (\ref{eq:GSa}) we can get the following differential equation for the scalar product $\langle\alpha|\psi\rangle$:
\begin{equation}\label{eq:DFG}
\begin{split}
    &\left[F\left(\alpha^*, \frac{\alpha}{2} + \frac{\partial}{\partial\alpha^*}\right) +
    i \lambda G\left(\alpha^*, \frac{\alpha}{2} + \frac{\partial}{\partial\alpha^*}\right)\right]
    \langle\alpha|\psi\rangle \\
    &= \beta \langle\alpha|\psi\rangle.
\end{split}
\end{equation}
According to Eq.~(\ref{eq:FBr}) we can look for the solution $\langle\alpha|\psi\rangle$ in the form
\begin{equation}
    \langle\alpha|\psi\rangle = \psi(\alpha^*) e^{-|\alpha|^2/2},
\end{equation}
whereby Eq.~(\ref{eq:DFG}) can be simplified to an ordinary differential equation of a complex variable
\begin{equation}\label{eq:DFG2}
    \left[F\left(\alpha^*, \frac{d}{d\alpha^*}\right) +
    i \lambda G\left(\alpha^*, \frac{d}{d\alpha^*}\right)\right]\psi(\alpha^*) =
    \beta \psi(\alpha^*).
\end{equation}
To get an ordinary differential equation of a real variable, let us represent $\alpha^*$ in polar coordinates as $\alpha^*=re^{-i\varphi}$. The unknown function $\psi(\alpha^*)$ can be considered as a function of the radius $r$ for a fixed phase $\varphi$
\begin{equation}\label{eq:tpr}
    \psi(\alpha^*) = \psi(r e^{-i \varphi}) = \psi_\varphi(r).
\end{equation}
The derivative $\psi^\prime_\varphi(r)$ (with respect to $r$) can be calculated with the help of the standard chain rule
\begin{equation}
    \psi^\prime_\varphi(r)= \frac{d \psi(\alpha^*)}{d \alpha^*} \frac{d \alpha^*}{d r}
    = \frac{d \psi(\alpha^*)}{d \alpha^*} e^{-i \varphi},
\end{equation}
thereby the derivative with respect to complex argument $\alpha^*$ is related to the derivative with respect to the radius (for the phase fixed) via
\begin{equation}
    \frac{d}{d \alpha^*} = e^{i \varphi} \frac{d}{d r}.
\end{equation}
Now Eq.~(\ref{eq:DFG}) can be written as an ordinary differential equation
\begin{equation}\label{eq:dt2}
\begin{split}
    &\left[F\left(r e^{-i \varphi}, e^{i \varphi}\frac{d}{d r}\right) +
    i \lambda G\left(r e^{-i \varphi}, e^{i \varphi}\frac{d}{d r}\right)\right] \psi_\varphi(r) \\
    &= \beta \psi_\varphi(r),
\end{split}
\end{equation}
or, more precisely, as a family of equations parameterized with the phase $\varphi \in [0, 2\pi]$. Solving this equation for all phases and combining the solutions $\psi_\varphi(r)$ together we get a solution $\psi(\alpha^*)$ of Eq.~(\ref{eq:DFG2}). The initial conditions to Eq.~(\ref{eq:dt2}) depend on the phase $\varphi$, but the final solution $\psi(\alpha^*)$ must be an analytical function of $\alpha^*$, thereby one cannot take  arbitrary functions of $\varphi$ as initial conditions to Eq.~(\ref{eq:dt2}). For $\psi(\alpha^*)$ to be analytical the initial conditions must be chosen as follows:
\begin{equation}\label{eq:GSic}
    \psi^{(k)}_\varphi(0) = c_k e^{-ik\varphi}, \quad k = 0, 1, \ldots,
\end{equation}
with $c_k$ being arbitrary complex numbers. The deficiency of this method is
that, in general, for arbitrarily chosen initial conditions (\ref{eq:GSic}),
the solution $\psi(\alpha^*)$ we get in this way is not normalized, one must
numerically take the integral (\ref{eq:FBn2}) to find the normalization
constant. In the figures of the numerically calculated $Q$-functions presented
below, we did not normalize the corresponding states, since in those figures
it is just a matter of scaling, which is unimportant for our purposes.

\section{Nonlinear quadrature squeezing}

In this section we introduce a nonlinear generalization of quadrature squeezed
states and we prove their nonclassicality.
First we define the nonlinear generalization of the quadratures to be
studied. The known special cases of quadrature squeezing and amplitude-squared
squeezing will be  reconsidered from this generalized point of view in the
following section.

%\subsection{Nonlinear Quadrature squeezing} 

Any pair of Hermitian operators $\hat{F}$ and $\hat{G}$ can be represented in the form
\begin{equation}\label{eq:Ef}
    \hat{F} = \hat{f} + \hat{f}^\dagger, \quad \hat{G} = -i(\hat{f}-\hat{f}^\dagger)
\end{equation}
for some operator $\hat{f}$, one may just set $\hat{f} = (\hat{F}+i\hat{G})/2$.
In all the examples considered below we use this representation with different choices of $\hat{f}$. 
Note that the operators that are defined in the following may be considered as a direct nonlinear generalization of the quadrature operators, which are recovered in the linear special case, $\hat{f}= \hat{a}$.  

Let us consider now the case of the operator $\hat{f}$ being a function of the annihilation operator only, $\hat{f} = f(\hat{a})$. In order to not overload the notation and without loss of generality, we assume in the following the function $f(z)$ to be real (i.e. for real argument $z$ its value $f(z)$ is also real), which implies that the relation $f(\hat{a})^\dagger = f(\hat{a}^\dagger)$ is valid. Let us write the generalized squeezing conditions given by the inequalities (\ref{eq:GSc}) explicitly. The first inequality is
\begin{equation}\label{eq:GAc}
    (\Delta F)^2 < \frac{1}{2} |\langle[\hat{F}, \hat{G}]\rangle|.
\end{equation}
The variance $(\Delta F)^2$ reads as
\begin{equation}
    (\Delta F)^2 = \langle(\Delta \hat{f}^\dagger)^2 + (\Delta \hat{f})^2 +
    \Delta \hat{f}^\dagger \Delta \hat{f} + \Delta \hat{f} \Delta \hat{f}^\dagger\rangle,
\end{equation}
and for the commutator of $\hat{F}$ and $\hat{G}$ we have the equality $\langle[\hat{F}, \hat{G}]\rangle = 2i\langle[\hat{f}, \hat{f}^\dagger]\rangle$. In \ref{app:NOa} it is shown that that the latter commutator is always nonnegative (in the case under study): $\langle[\hat{f}, \hat{f}^\dagger]\rangle = \langle[\Delta\hat{f}, \Delta\hat{f}^\dagger]\rangle \geq 0$. Now the inequality (\ref{eq:GAc}) can be written as follows:
\begin{equation}\label{eq:in1}
    \langle(\Delta \hat{f}^\dagger)^2\rangle + \langle(\Delta \hat{f})^2\rangle +
    2\langle\Delta \hat{f}^\dagger \Delta \hat{f}\rangle < 0.
\end{equation}
Since $\hat{f} = f(\hat{a})$, the left-hand side of this inequality is just $\langle:(\Delta \hat{F})^2:\rangle$, so the condition (\ref{eq:GAc}) reads as
\begin{equation}
    \langle:(\Delta \hat{F})^2:\rangle < 0.
\end{equation}
From the second inequality of (\ref{eq:GSc}) we can conclude in the same way that $\langle:(\Delta \hat{G})^2:\rangle < 0$. We see that generalized quadrature squeezing always implies nonclassical behavior. 

The eigenvalue problem (\ref{eq:GSevp}) reads as
\begin{equation}\label{eq:fa}
    \bigl((1+\lambda)f(\hat{a}) + (1-\lambda)f(\hat{a}^\dagger)\bigr) |\psi\rangle =
    \beta |\psi\rangle,
\end{equation}
and the corresponding differential equation (\ref{eq:dt2}) as
\begin{equation}\label{eq:fad}
    f\left(e^{i\varphi}\frac{d}{dr}\right)\psi_\varphi(r) =
    \left(\frac{\lambda-1}{\lambda+1}f(re^{-i\varphi})+\frac{\beta}{\lambda+1}\right)
    \psi_\varphi(r).
\end{equation}
In general, for an arbitrary function $f(z)$, it is impossible to solve these equations analytically. But in a special case of $\lambda=1$ it is possible to find the general solution, provided that we know how to find the roots of entire functions. In the case of $\lambda=1$ Eq.~(\ref{eq:fa}) simply reads as $f(\hat{a})|\psi\rangle = \gamma|\psi\rangle$, where $\gamma = \beta/2$. Each root $\alpha$ to the equation $f(\alpha)=\gamma$ gives a partial solution of the form $P_{k_\alpha-1}(\hat{a}^\dagger-\alpha)|\alpha\rangle$, where $k_\alpha$ is the multiplicity of the root $\alpha$ and $P_{k_\alpha-1}(z)$ is an arbitrary polynomial of the degree $k_\alpha-1$. In particular, any simple root (of multiplicity $1$) gives the solution which is proportional to the coherent state $|\alpha\rangle$. The general solution $|\psi\rangle$ of Eq.~(\ref{eq:fa}) for $\lambda=1$ is a linear combination of all the partial solutions
\begin{equation}\label{eq:gs}
    |\psi\rangle = \sum_{f(\alpha) = \gamma} P_{k_\alpha-1}(\hat{a}^\dagger - \alpha^*)
    |\alpha\rangle,
\end{equation}
where the sum here is taken over all the roots of the equation $f(\alpha)=\gamma$. If all the roots of this equation are simple then the general solution (\ref{eq:gs}) is just a linear combination of coherent states. Below we consider only polynomial functions $f(z)$. Note that, though there is no general analytical expression for the roots of a polynomial of a degree greater than four, it is possible to find out whether all the roots of the polynomial in question are simple or not (a polynomial whose all roots are simple is called separable), which answers the question whether the sum (\ref{eq:gs}) contains only coherent states or not. A polynomial is separable if and only if a special determinant constructed from the coefficients of the polynomial is not equal to zero. The details see, for example, in \cite{cox}.

\begin{figure*}
\begin{center}
    \subfigure[$\lambda=0.1$]{\includegraphics[scale=0.28]{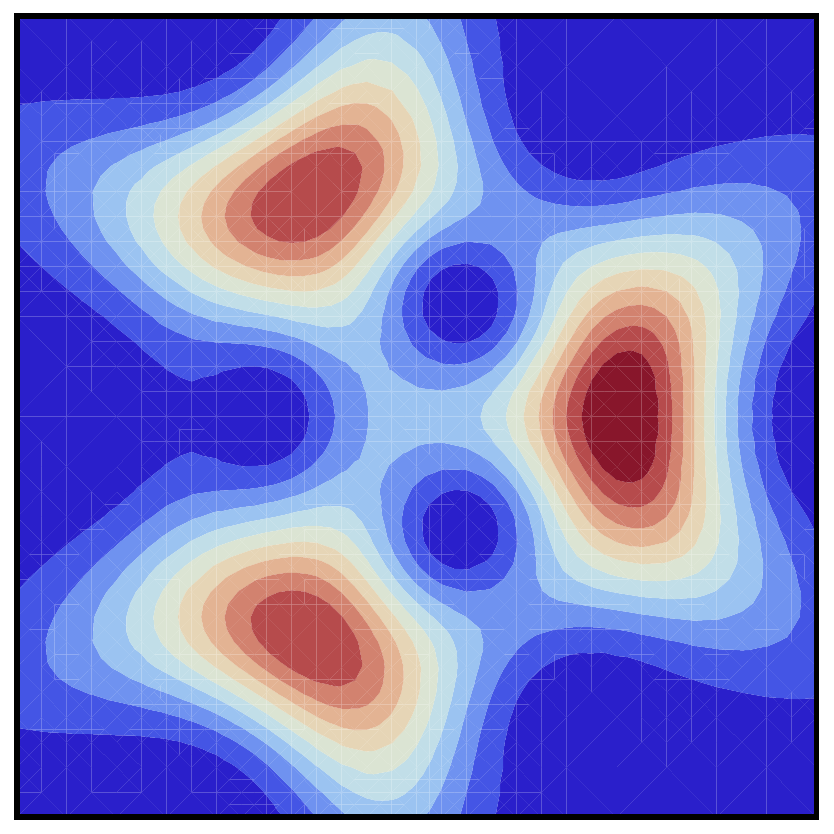}}
    \subfigure[$\lambda=0.2$]{\includegraphics[scale=0.28]{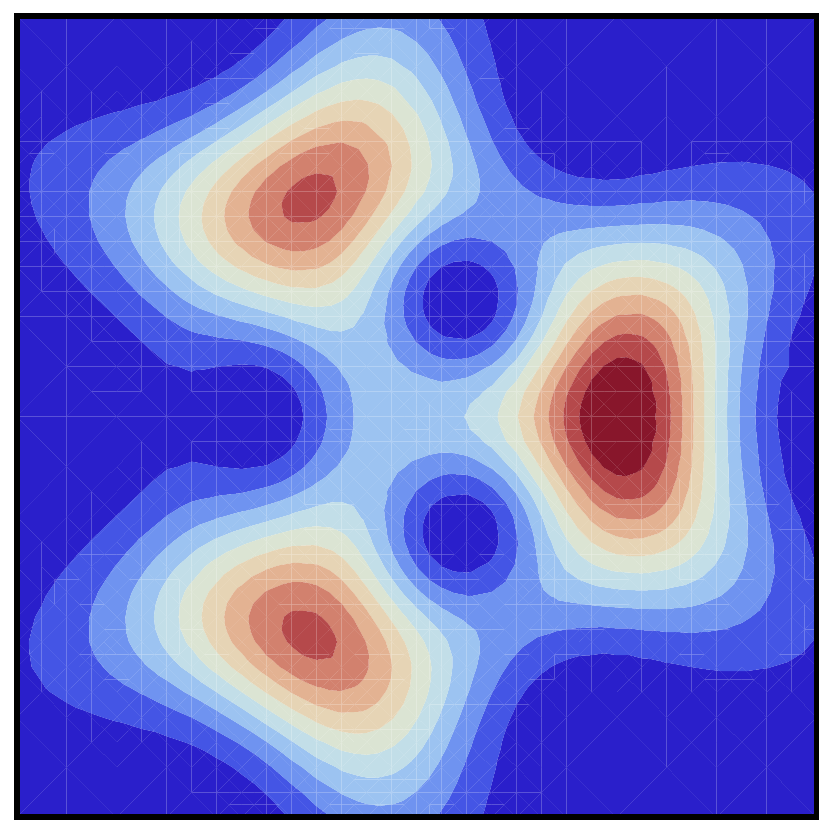}}
    \subfigure[$\lambda=0.5$]{\includegraphics[scale=0.28]{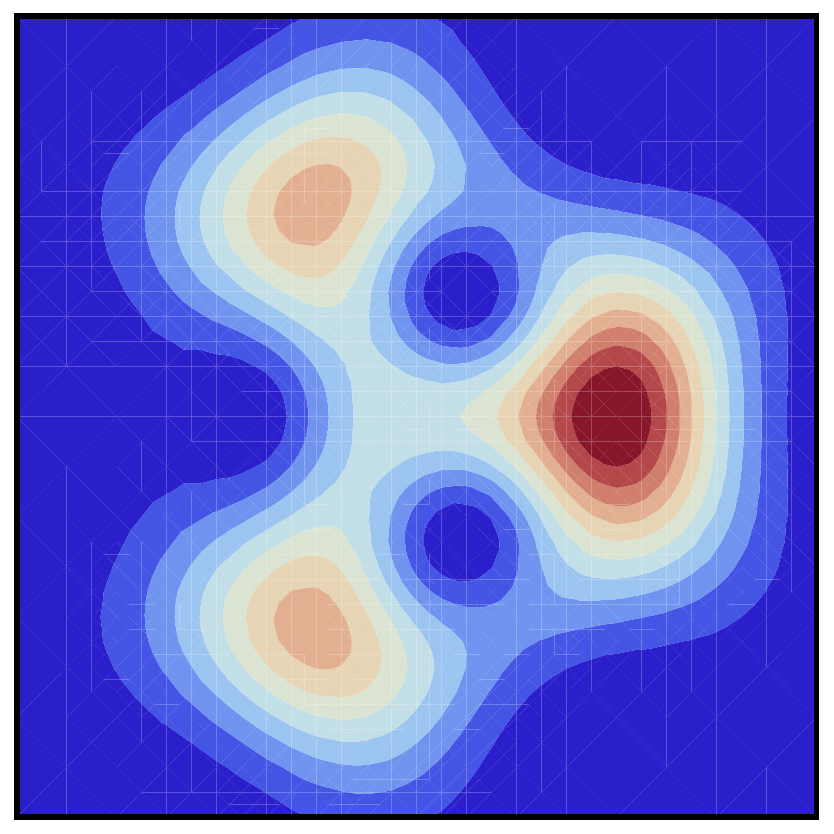}}
    \subfigure[$\lambda=1$]{\includegraphics[scale=0.28]{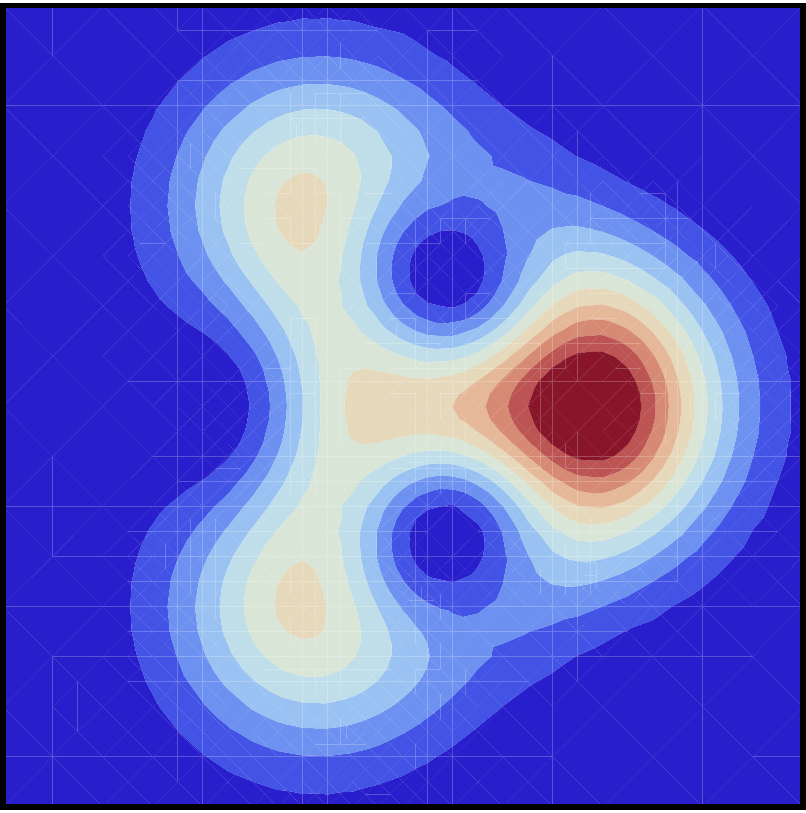}}
    \subfigure[$\lambda=2$]{\includegraphics[scale=0.28]{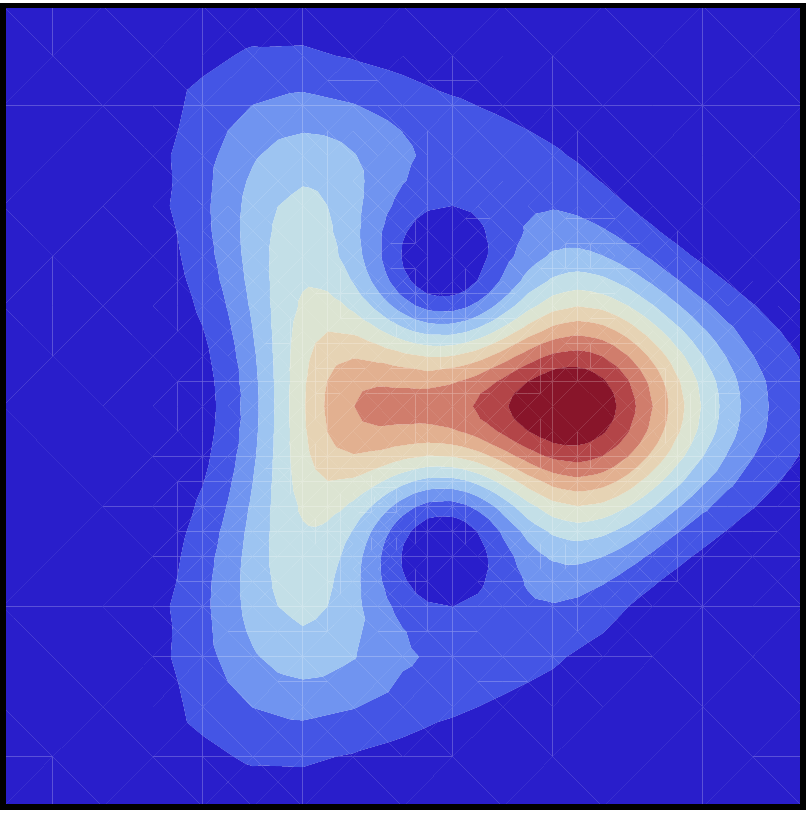}}
    \subfigure[$\lambda=5$]{\includegraphics[scale=0.28]{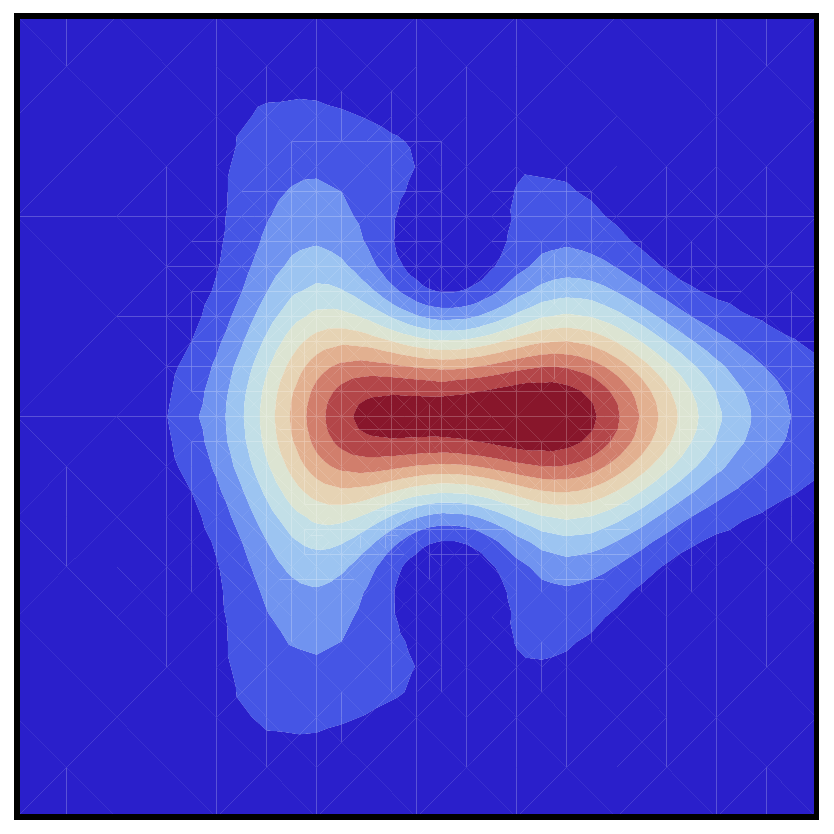}}
    \subfigure[$\lambda=10$]{\includegraphics[scale=0.28]{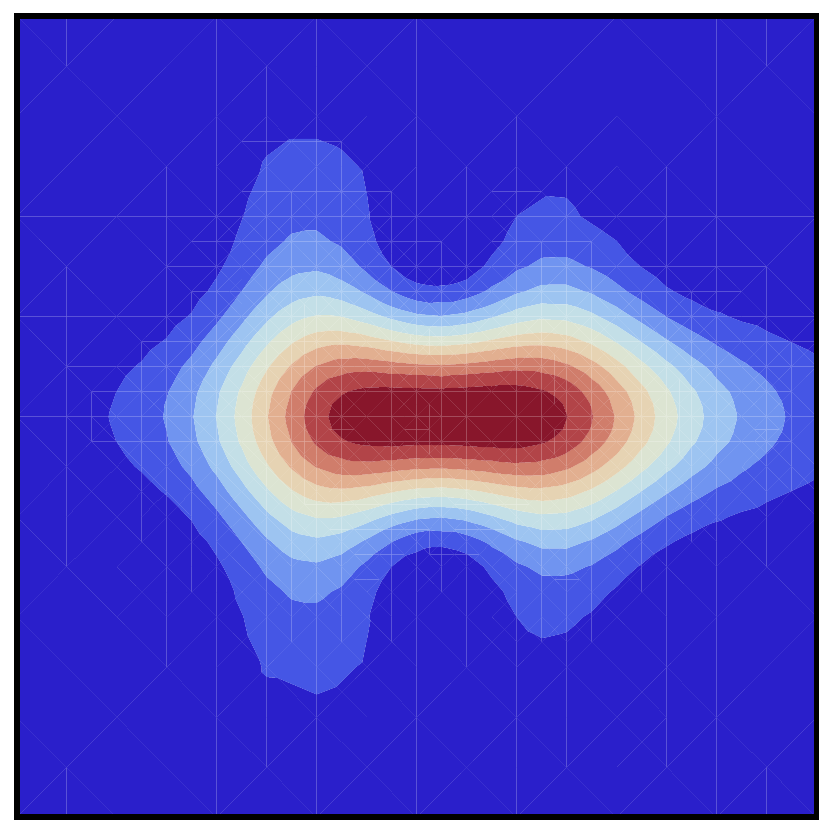}}
\end{center}
\caption{The $Q$-function corresponding to the solution of Eq.~(\ref{eq:f1}) with the initial conditions $\theta_\varphi(0) = 1$, $\theta^\prime_\varphi(0) = 0$, $\theta^{\prime\prime}_\varphi(0) = e^{-2i\varphi}$.}\label{fig:f1}
\end{figure*}

It is clear to see that Eq.~(\ref{eq:fad}) with a first-order order polynomial $f(z)$ corresponds to the definition of quadrature squeezing, and it is easy to prove that a second-order polynomial leads to the definition of amplitude-squared squeezing. That is why it makes sense to consider polynomials of third and higher orders. As an example, let us consider the polynomial $f(z) = z^3+z$. The differential equation (\ref{eq:fad}) with this polynomial reads as
\begin{equation}\label{eq:f1}
\begin{split}
    &e^{3 i \varphi} \psi^{\prime\prime\prime}_\varphi(r) + e^{i \varphi}
    \psi^{\prime}_\varphi(r) \\
    &= \left(\frac{\lambda-1}{\lambda+1}(r^3 e^{-3 i \varphi} + r
    e^{- i \varphi})+\frac{\beta}{\lambda+1}\right) \psi_\varphi(r).
\end{split}
\end{equation}
The contour plot of the $Q$-function of a solution of this equation is shown in Fig.~\ref{fig:f1} for some values of the squeezing parameter $\lambda$. 

In the next sections we consider the two cases (quadrature squeezing and amplitude-squared squeezing) in a more general context, allowing the squeezing parameter $\lambda$ to be complex. We find the condition which guarantees that solutions of the corresponding equations are normalizable (and thus can be identified with physical quantum states) and show that only for real $\lambda$ are these states minimum uncertainty. The details of the calculation are given in the appendices.

\section{Quadrature and amplitude-squared squeezing revisited}

Let us now reconsider the well-known cases of quadrature squeezing and amplitude-squared squeezing in the framework of our approach.
For these cases we can solve the problems analytically. In the next section we
will further develop our approach for the study of deformed nonlinear squeezing. 

\subsection{Quadrature squeezing}

In the case of quadrature squeezing the operator $\hat{f}$ is simply the annihilation operator $\hat{f}=\hat{a}$. The operators $\hat{F}$ and $\hat{G}$ defined by Eq.~(\ref{eq:Ef}) are then two orthogonal quadratures: $\hat{F}=\hat{x}$ and $\hat{G}=\hat{p}$. The differential equation (\ref{eq:DFG2}) in this case reads as
\begin{equation}
    \frac{d\psi(\alpha^*)}{d\alpha^*} = \left(\frac{\lambda-1}{\lambda+1}\alpha^*+\frac{\beta}{\lambda+1}\right) \psi(\alpha^*).
\end{equation}
This equation can be easily solved which gives us the eigenstates of the corresponding eigenvalue problem (\ref{eq:fa})
\begin{equation}\label{eq:psi1}
    |\psi\rangle = \mathcal{N} \exp\left(\frac{1}{2}\frac{\lambda-1}{\lambda+1}\hat{a}^{\dagger 2}+\frac{\beta}{\lambda+1}\hat{a}^\dagger\right)|0\rangle.
\end{equation}
To calculate the normalization $\mathcal{N}$ using the relation (\ref{eq:no}) we need the following equality for Hermite polynomials:
\begin{equation}\label{eq:hermite}
\begin{split}
    \sum^{+\infty}_{n=0} & H_n(x) H_n(y) \frac{(z/2)^n}{n!} \\
    &= \frac{1}{\sqrt{1-z^2}}\exp\left(\frac{2xyz-(x^2+y^2)z^2}{1-z^2}\right).
\end{split}
\end{equation}
The series on the left hand side converges only if $|z|<1$. Here for $x$, $y$ and $z$ we get
\begin{equation}
    x = \frac{i}{\sqrt{2}}\frac{\beta}{\sqrt{\lambda^2-1}}, \quad y = x^*, \quad
    z = \left|\frac{\lambda-1}{\lambda+1}\right|.
\end{equation}
The condition of convergence $|z|<1$ is equivalent to the positivity of $\re\lambda$, thereby the state (\ref{eq:psi1}) is normalizable if and only if $\re\lambda>0$. The normalization $\mathcal{N}$ is given by
\begin{equation}
    \mathcal{N}^2 = \frac{2\sqrt{\re\lambda}}{|\lambda+1|}
    \exp\left(-\frac{|\beta|^2+\re\left(\frac{\lambda^*-1}{\lambda+1}\beta^2\right)}
    {4\re\lambda}\right).
\end{equation}
For $\lambda=1$ the state (\ref{eq:psi1}) is just a coherent state $|\beta/2\rangle$.

For real $\lambda$ the state (\ref{eq:psi1}) minimizes the uncertainty relation (\ref{eq:GSur}): $\Delta x \Delta p = 1$. Let us see what happens when $\lambda$ is complex. To find the dispersions of $\hat{x}$ and $\hat{p}$ for the state (\ref{eq:psi1}) we need to calculate its normally ordered moments up to second order. Note that it is straightforward to calculate the antinormally ordered moments $\langle \hat{a}^n \hat{a}^{\dagger m}\rangle$ for the state (\ref{eq:psi1}) due to the following relation:
\begin{equation}\label{eq:QSam}
    \langle \hat{a}^n \hat{a}^{\dagger m}\rangle = \mathcal{N}^2 (\lambda^*+1)^n(\lambda+1)^m \frac{\partial^{n+m} \mathcal{N}^{-2}}{\partial\beta^{*n} \partial\beta^m}.
\end{equation}
It is possible to express these moments in terms of Hermite polynomials and then obtain normally ordered moments in general, but we need only second-order moments, which can be easily obtained directly from Eq.~(\ref{eq:QSam}). For the simplest moment $\langle\hat{a}\rangle$ we have
\begin{equation}
    \langle\hat{a}\rangle = \frac{\re(\lambda\beta^*)+i\im\beta}{2\re\lambda},
\end{equation}
The expressions for $\langle\hat{a}^2\rangle$ and $\langle\hat{a}^\dagger \hat{a}\rangle$ are more lengthy, so we present here only the final expressions for the dispersions
\begin{equation}
    (\Delta x)^2 = \frac{|\lambda|^2}{\re\lambda}, \quad 
    (\Delta p)^2 = \frac{1}{\re\lambda}.
\end{equation}
To see how much the product of the dispersions differs from the lowest possible value $1$ let us calculate the difference
\begin{equation}\label{eq:QSd}
    (\Delta x)^2(\Delta p)^2 - 1 = \frac{|\lambda|^2}{\re^2\lambda} - 1 = \tan^2\Phi,
\end{equation}
where $\Phi = \arg\lambda$. We see that the state (\ref{eq:psi1}) is minimum uncertainty if and only if $\lambda$ is real, and the larger $\Phi$ is, the stronger the equality $\Delta x \Delta p = 1$ is violated. This is closely related to the problem of optimal choice of the phase of the quadratures adjusted to the principal axes of the squeezing ellipses, cf. \cite{pla-68-321}. When $\lambda$ tends to the imaginary axis (i.e. when $\varphi$ tends to $\pm\pi/2$), the difference (\ref{eq:QSd}) tends to infinity.

It is also interesting to see how the squeezing of the state (\ref{eq:psi1}) depends on $\lambda$. To see this we must calculate the minimum $\min_\varphi \langle:(\Delta \hat{x}_\varphi)^2:\rangle$ of the general quadrature $\hat{x}_\varphi = \hat{a}e^{-i\varphi} + \hat{a}^\dagger e^{i\varphi}$. It is not difficult to do and after all simplifications this minimum reads as follows:
\begin{equation}
    \min_\varphi \langle:(\Delta \hat{x}_\varphi)^2:\rangle = 
    -\frac{2|\lambda-1|}{|\lambda+1|+|\lambda-1|}.
\end{equation}
We see that the state under study is always squeezed, except the case of $\lambda=1$, when it is just a coherent state, as it has already been mentioned above.

\subsection{Amplitude-squared squeezing}

\begin{figure*}
\begin{center}
    \subfigure[$\lambda=0.1+i$]{\includegraphics[scale=0.28]{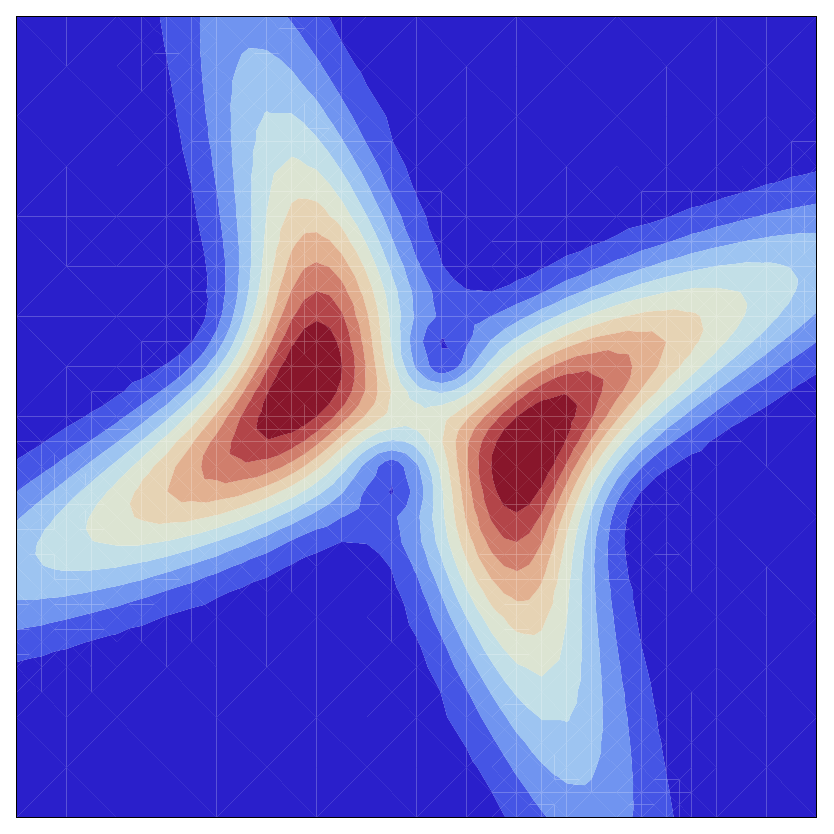}}
    \subfigure[$\lambda=0.2+i$]{\includegraphics[scale=0.28]{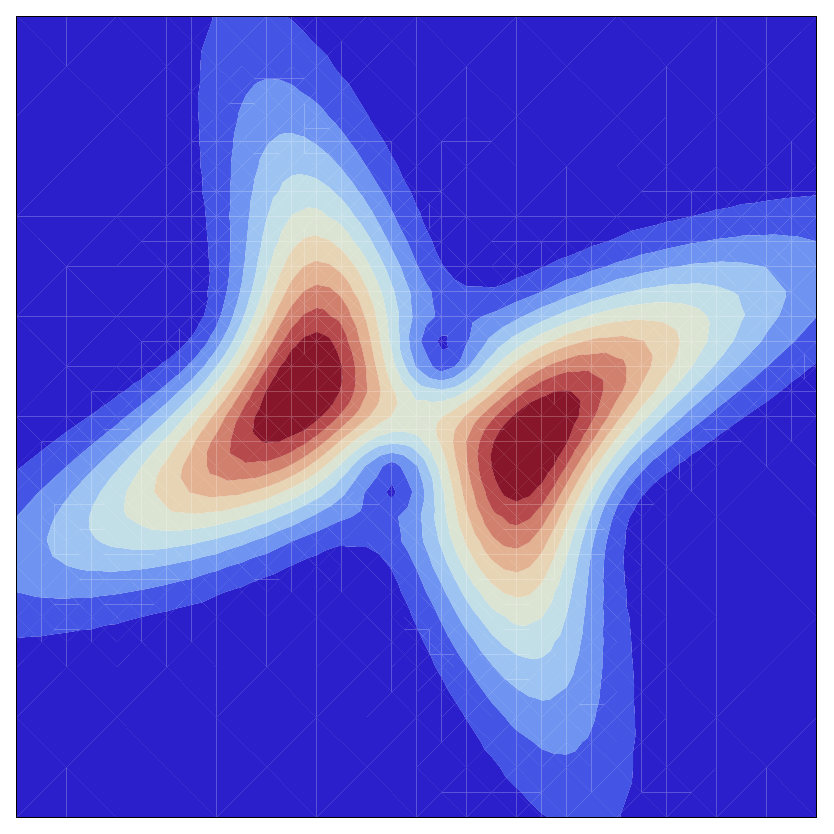}}
    \subfigure[$\lambda=0.5+i$]{\includegraphics[scale=0.28]{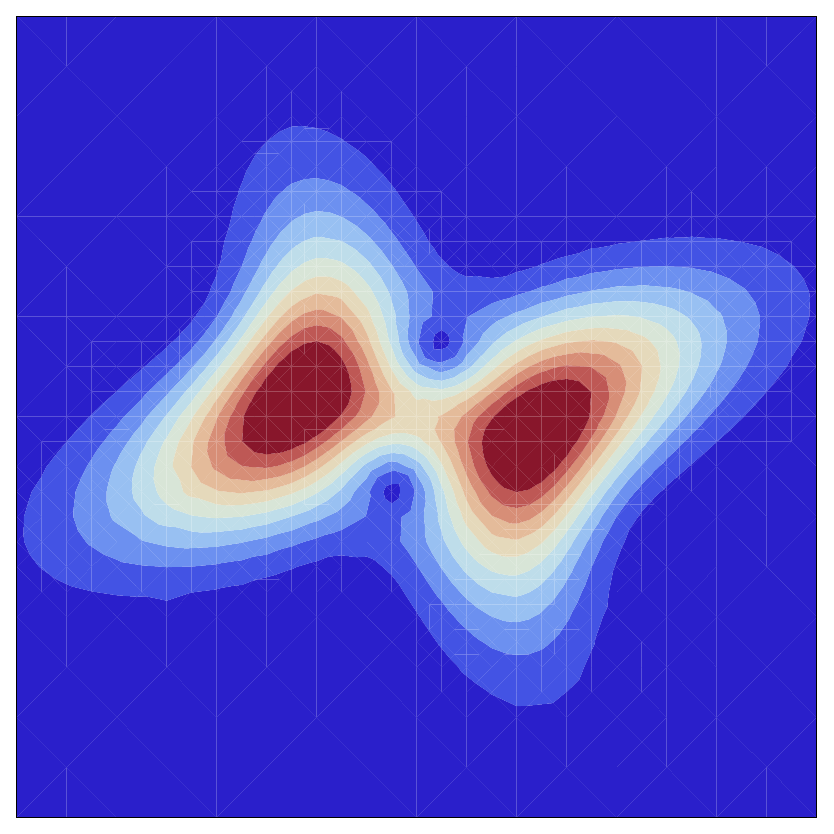}}
    \subfigure[$\lambda=1+i$]{\includegraphics[scale=0.28]{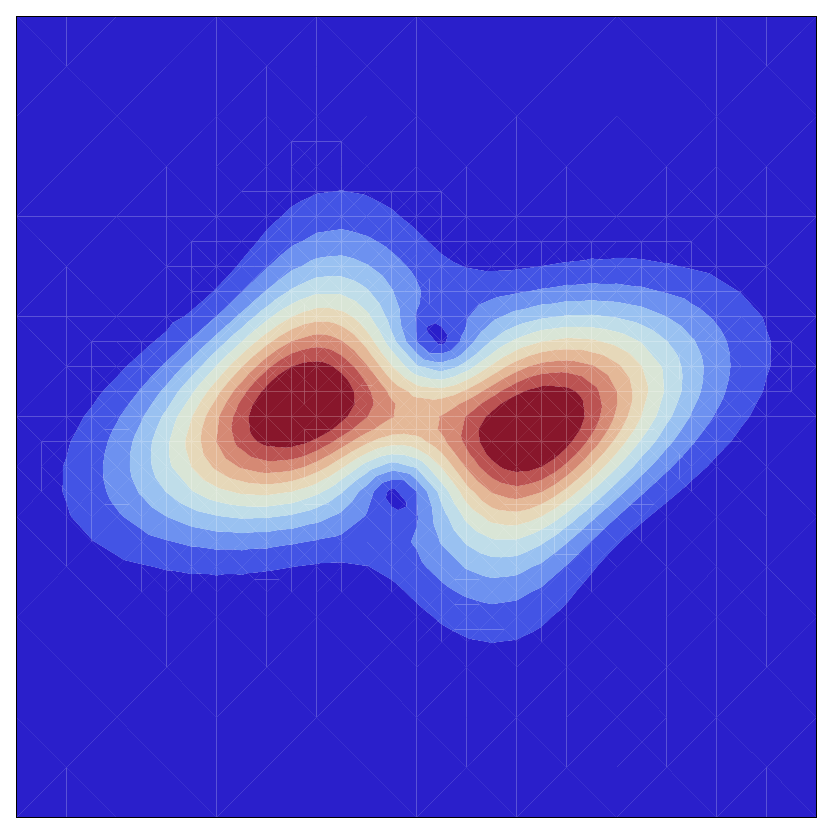}}
    \subfigure[$\lambda=2+i$]{\includegraphics[scale=0.28]{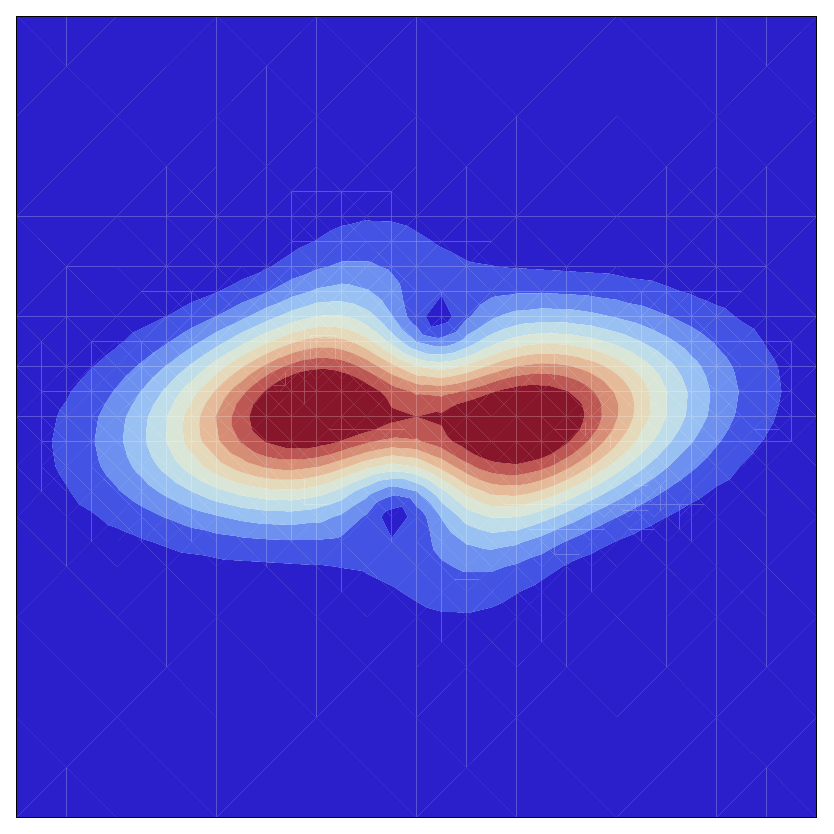}}
    \subfigure[$\lambda=5+i$]{\includegraphics[scale=0.28]{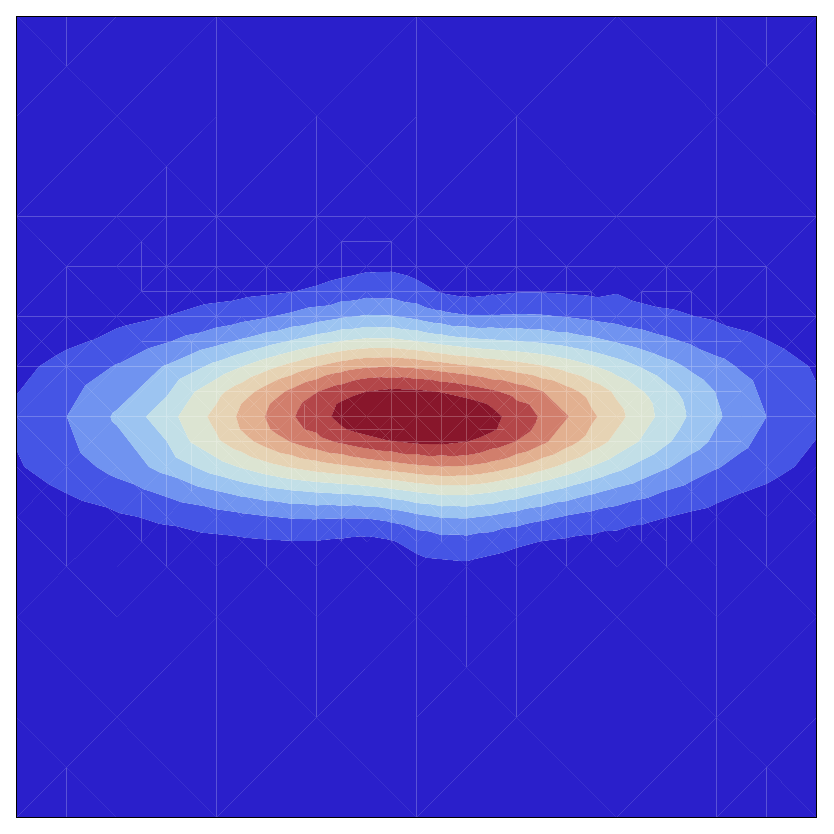}}
    \subfigure[$\lambda=10+i$]{\includegraphics[scale=0.28]{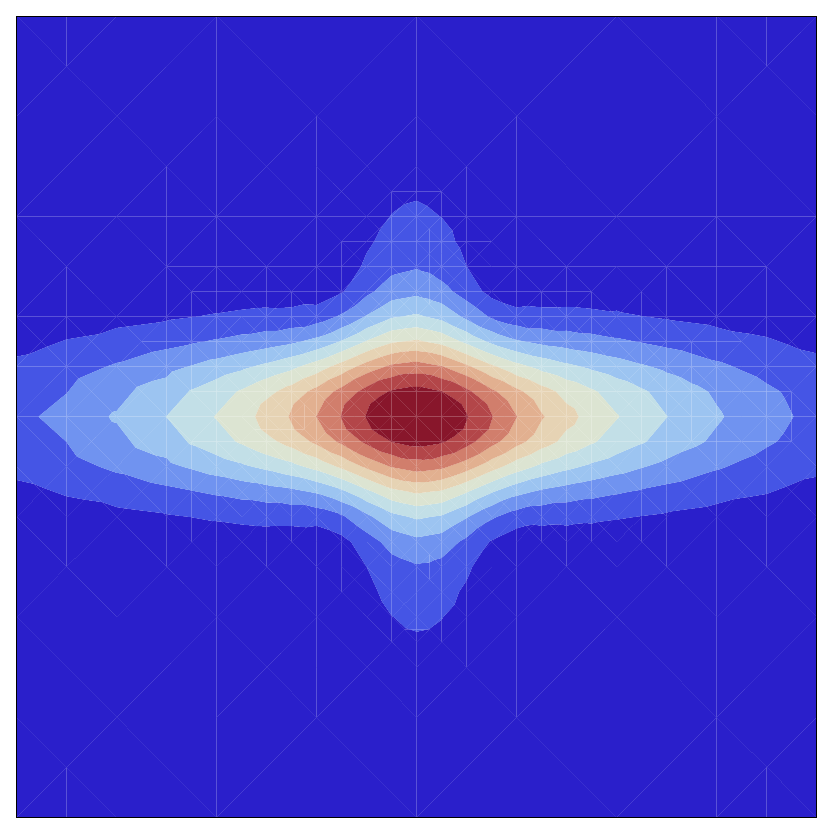}}
\end{center}
\caption{The $Q$-function of the even state (\ref{eq:ASp}) for $\beta=5$ and different $\lambda$.}\label{fig:f2}
\end{figure*}

In the case of amplitude squared squeezing the operator $\hat{f}$ reads as $\hat{f}=\hat{a}^2$. The differential equation (\ref{eq:DFG2}) is now a second-order equation
\begin{equation}
    \frac{d^2\psi(\alpha^*)}{d\alpha^*} - \left(\frac{\lambda-1}{\lambda+1}\alpha^{* 2}+\frac{\beta}{\lambda+1}\right)\psi(\alpha^*) = 0.
\end{equation}
It has the following two linearly independent solutions (called even and odd, for self-evident reason):
\begin{equation}\label{eq:ASs}
\begin{split}
    \psi_{\mathrm{e}}(b; c; \alpha^*) &= e^{-c \alpha^{*2}} {}_1F_1\left(b; \frac{1}{2}; 2c\alpha^{*2}\right), \\ 
    \psi_{\mathrm{o}}(b; c; \alpha^*) &= \alpha^* e^{-c \alpha^{*2}} {}_1F_1\left(b+\frac{1}{2}; \frac{3}{2}; 2c\alpha^{*2}\right),
\end{split}
\end{equation}
where ${}_1F_1(b; c; z)$ is the Kummer function defined via
\begin{equation}\label{eq:kummer}
    {}_1F_1(b; c; z) = \sum^{+\infty}_{k=0} \frac{(b)_k}{(c)_k} \frac{z^k}{k!},
\end{equation}
with $(b)_k$ being the Pochhammer symbol expressed in terms of gamma functions as $(b)_k = \Gamma(b+k)/\Gamma(b) = b(b+1) \ldots (b+k-1)$. The parameters $b$ and $c$ read as
\begin{equation}\label{eq:bc}
    c = \frac{1}{2} \sqrt{\frac{\lambda-1}{\lambda+1}}, \quad
    b = \frac{1}{4} \left(1+\frac{\beta}{\sqrt{\lambda^2-1}}\right).
\end{equation}
Note that there are two choices for the parameter $c$ which differ in sign only (since each nonzero complex number has two square roots). When the sign of the square root $\sqrt{(\lambda-1)/(\lambda+1)}$ is chosen, the square root $\sqrt{\lambda^2-1}$ for the parameter $b$ must be calculated according to the equality $\sqrt{\lambda^2-1} = (\lambda-1)/\sqrt{(\lambda-1)/(\lambda+1)}$. In \ref{app:ASS} it is shown that both the solutions $\psi_{\mathrm{e}}$ and $\psi_{\mathrm{o}}$, Eq.~(\ref{eq:ASs}), do not depend on the choice of the sign of the square root for $c$.

It is straightforward to write the quantum states which correspond to the functions (\ref{eq:ASs}) in the Fock-Bargmann representation, but for us it is more convenient to represent them in another form. In fact, as shown in \ref{app:ASS}, the even and odd solutions of the eigenvalue problem (\ref{eq:fa}) read as
\begin{equation}\label{eq:ASp}
\begin{split}
    |\psi_{\mathrm{e}}\rangle &= \mathcal{N}_{\mathrm{e}} S(\xi) {}_1F_1\left(b; \frac{1}{2}; v \hat{a}^{\dagger 2}\right) |0\rangle, \\
    |\psi_{\mathrm{o}}\rangle &= \mathcal{N}_{\mathrm{o}} S(\xi) {}_1F_1\left(b+\frac{1}{2}; \frac{3}{2}; v \hat{a}^{\dagger 2}\right) |1\rangle,
\end{split}
\end{equation}
where the parameters $\xi$ and $v$ are defined via
\begin{equation}\label{eq:ASxv}
    e^{i\arg\xi} \tanh|\xi| = \sqrt{\frac{\lambda-1}{\lambda+1}}, \quad
    v = \frac{\sqrt{\frac{\lambda-1}{\lambda+1}}}{1+\left|\frac{\lambda-1}{\lambda+1}\right|},
\end{equation}
and the normalizations $\mathcal{N}_{\mathrm{e}}$ and $\mathcal{N}_{\mathrm{o}}$ are as follows:
\begin{equation}\label{eq:ASn}
\begin{split}
    \mathcal{N}^{-2}_{\mathrm{e}} &= {}_2F_1\left(b, b^*; \frac{1}{2}; 4|v|^2\right), \\
    \mathcal{N}^{-2}_{\mathrm{o}} &= {}_2F_1\left(b+\frac{1}{2}, b^*+\frac{1}{2}; \frac{3}{2}; 4|v|^2\right).
\end{split}
\end{equation}
Note that for real $\lambda$ the states (\ref{eq:ASp}) are exactly the ones obtained in \cite{qo-6-37}. The contour plot of the $Q$-function of the even state is shown in Fig.~\ref{fig:f2}.

In the case of quadrature squeezing there are relatively compact analytical expressions for the general normally ordered moments of the solutions of the corresponding eigenvalue problem (\ref{eq:fa}). In the present case the situation is much more complicated. To our knowledge, there is no analytical expression for the characteristic function of the states (\ref{eq:ASp}), but using Eqs.~(\ref{eq:FBmc}) and (\ref{eq:FBm}) together with Eq.~(\ref{eq:FBsc}) it is possible to find analytical expressions for normally-ordered moments of these states. But these expressions are really huge, for example, only the single moment $\langle\hat{a}^4\rangle$ (which is needed to calculate the dispersions of $\hat{F}$ and $\hat{G}$) expands to a whopping whole-page expression. Let us present here only the mean photon number of the even state
\begin{equation}
\begin{split}
    \langle\hat{n}\rangle &= 4|b|^2 \frac{|\lambda^2-1|}{\re\lambda} \left(\frac{|\lambda+1|-|\lambda-1|}{|\lambda+1|+|\lambda-1|}\right)^2 \\
    &\times \frac{{}_2F_1\left(b+1, b^*+1; \frac{3}{2}; 4|c|^2\right)}{{}_2F_1\left(b, b^*; \frac{1}{2}; 4|c|^2\right)} \\
    &- 2\frac{|\lambda^2-1|}{\re\lambda} \re b + \frac{|\lambda-1|}{|\lambda+1|-|\lambda-1|}.
\end{split}
\end{equation}
Note again, that for real $\lambda$ this coincides with the expression for the mean photon number obtained in \cite{qo-6-37}. We found that the states (\ref{eq:ASp}) are minimum uncertainty only for real $\lambda$, and the larger the phase of $\lambda$ the stronger is the violation of the corresponding equation (\ref{eq:GSu}), though in this case the strength of the violation does not depend only on the phase of $\lambda$, as it was in the case of quadrature squeezing. Calculating the moments with \textsl{Mathematica} we experienced dramatic loss of accuracy working with machine precision, so it was necessary to use high-precision numbers to get the correct results.

\section{Deformed nonlinear squeezed states}

Here we consider a more complex case of the operator $\hat{f}$ from Eq.~(\ref{eq:Ef}) to be of the form $\hat{f} = g(\hat{n})\hat{a}$, where we again assume $g(z)$ to be real. The eigenvalue problem (\ref{eq:GSevp}) in this case reads as
\begin{equation}\label{eq:DSe}
    \Bigl((1+\lambda) g(\hat{n}) \hat{a} + (1-\lambda) \hat{a}^\dagger g(\hat{n})\Bigr)
    |\psi\rangle = \beta |\psi\rangle.
\end{equation}
To formulate the corresponding differential equation (\ref{eq:dt2}) we need to normally order the functions of the photon number operator. In \ref{app:NO} we prove the following relation:
\begin{equation}\label{eq:fne}
    g(\hat{n}) = \sum^{+\infty}_{k=0} \frac{(\Delta^k g)(0)}{k!} :\hat{n}^k:\ =\ :(e^{\hat{n} \Delta} g)(0):,
\end{equation}
where $\Delta$ is the difference operator defined via $(\Delta g)(n) = g(n+1)-g(n)$. The powers of this operator calculated at zero read as
\begin{equation}
    (\Delta^k g)(0) = \sum^k_{i=0} (-1)^{k-i} \binom{k}{i} g(i).
\end{equation}
Now we can write the differential equation (\ref{eq:dt2}) explicitly
\begin{equation}\label{eq:DSd}
    \sum^{+\infty}_{k=0} C_k(r) \frac{d^k \psi_\varphi(r)}{dr^k} = \beta \psi_\varphi(r),
\end{equation}
where the functions $C_k(r) \equiv C_k(\lambda, r, \varphi)$, $k \geq 1$, are defined via
\begin{equation}\label{eq:DSC}
\begin{split}
    C_0(r) &= (1-\lambda) g(0) r e^{-i \varphi},  \\
    C_k(r) &= (1+\lambda) \frac{(\Delta^{k-1}g)(0)}{(k-1)!} r^{k-1} e^{i \varphi} \\ 
           &+ (1-\lambda) \frac{(\Delta^k g)(0)}{k!} r^{k+1} e^{-i \varphi}.
\end{split}
\end{equation}
The initial conditions are given as in Eq.~(\ref{eq:GSic}), but in this case the coefficients of the equation are not constants and that is why the initial conditions cannot be chosen arbitrarily. It is shown in \ref{app:IC} that there are the following relations for the initial conditions to Eq.~(\ref{eq:DSd}), $k \geq 0$:
\begin{equation}\label{eq:nic}
\begin{split}
    k(1-\lambda) &g(k-1) e^{-i \varphi} \psi^{(k-1)}_\varphi(0) - \beta
    \psi^{(k)}_\varphi(0) \\
    &+ (1+\lambda) g(k) e^{i \varphi} \psi^{(k+1)}_\varphi(0) = 0.
\end{split}
\end{equation}
Usually (but not always, as we will see shortly) this means that the solution of Eq.~(\ref{eq:DSd}) is unique.

\begin{figure*}
\begin{center}
    \subfigure[$\lambda=0.1$]{\includegraphics[scale=0.28]{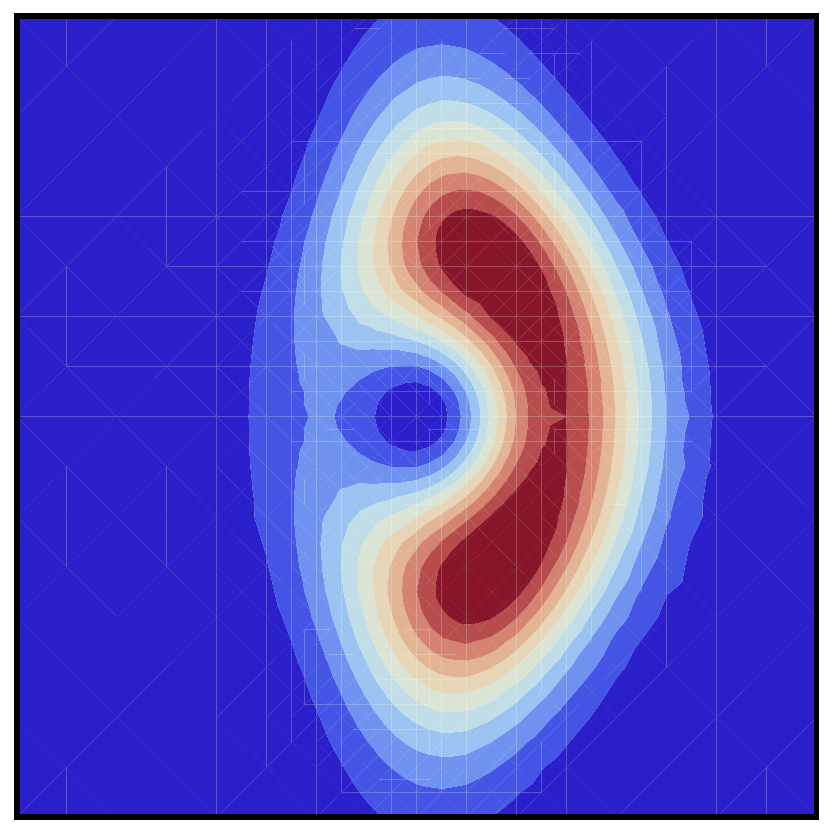}}
    \subfigure[$\lambda=0.2$]{\includegraphics[scale=0.28]{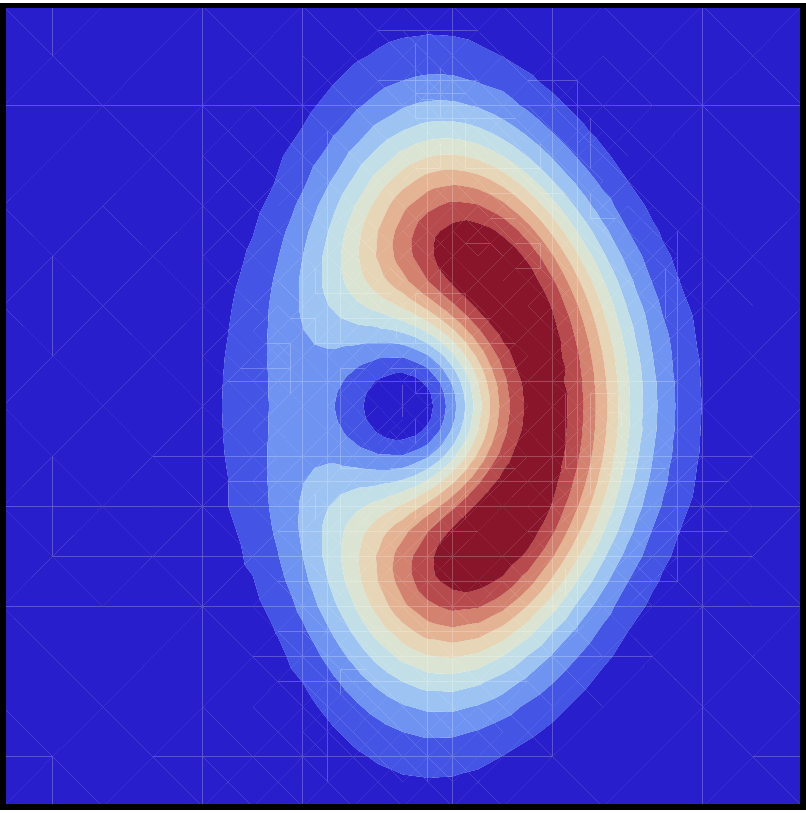}}
    \subfigure[$\lambda=0.5$]{\includegraphics[scale=0.28]{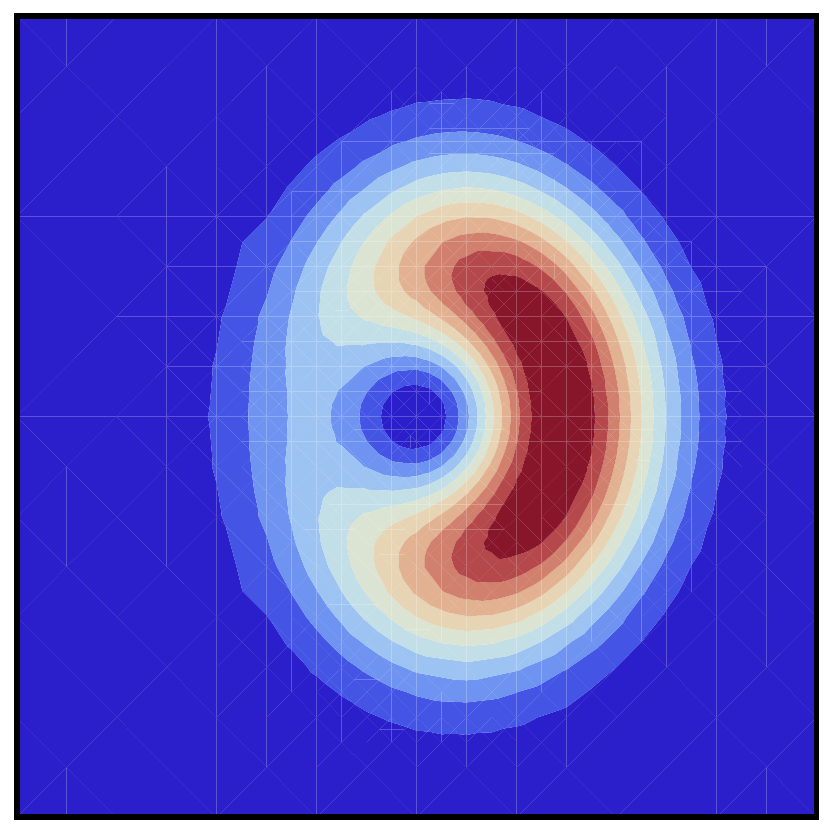}}
    \subfigure[$\lambda=1$]{\includegraphics[scale=0.28]{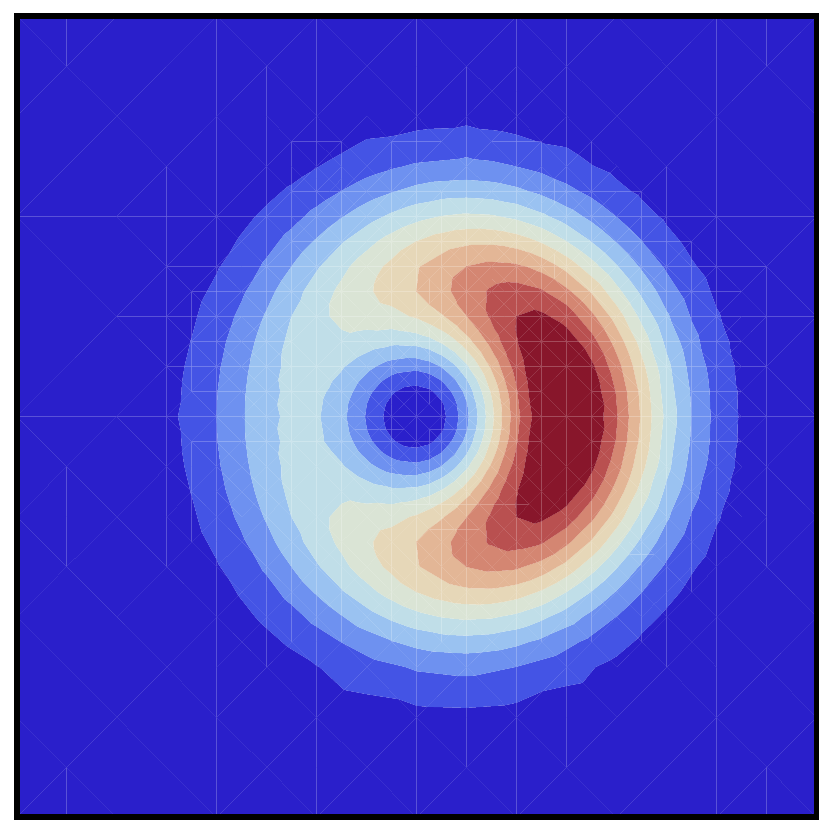}}
    \subfigure[$\lambda=2$]{\includegraphics[scale=0.28]{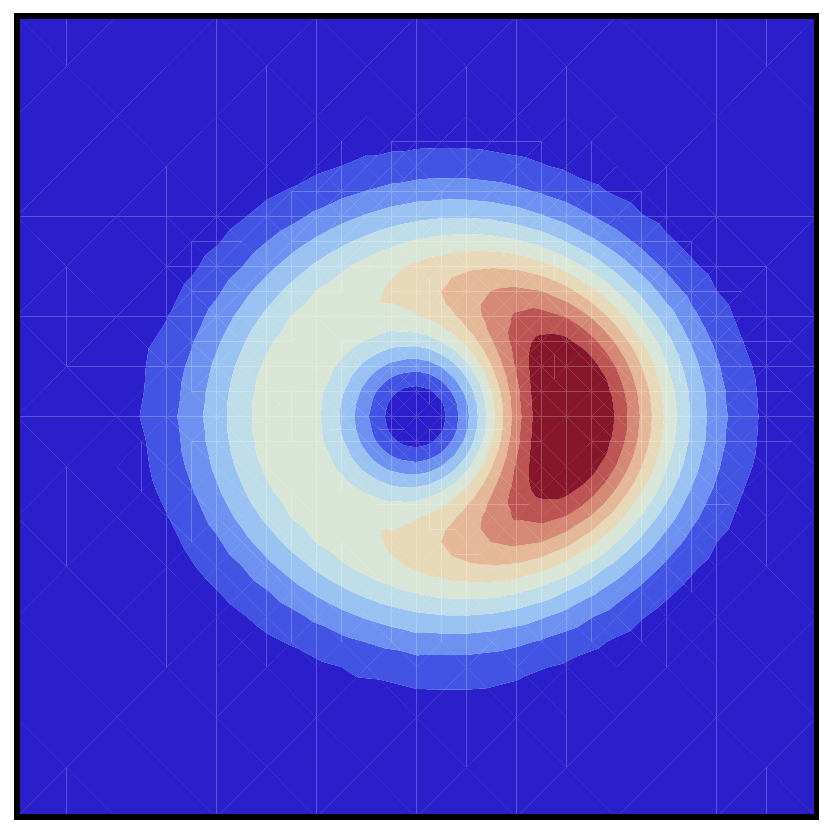}}
    \subfigure[$\lambda=5$]{\includegraphics[scale=0.28]{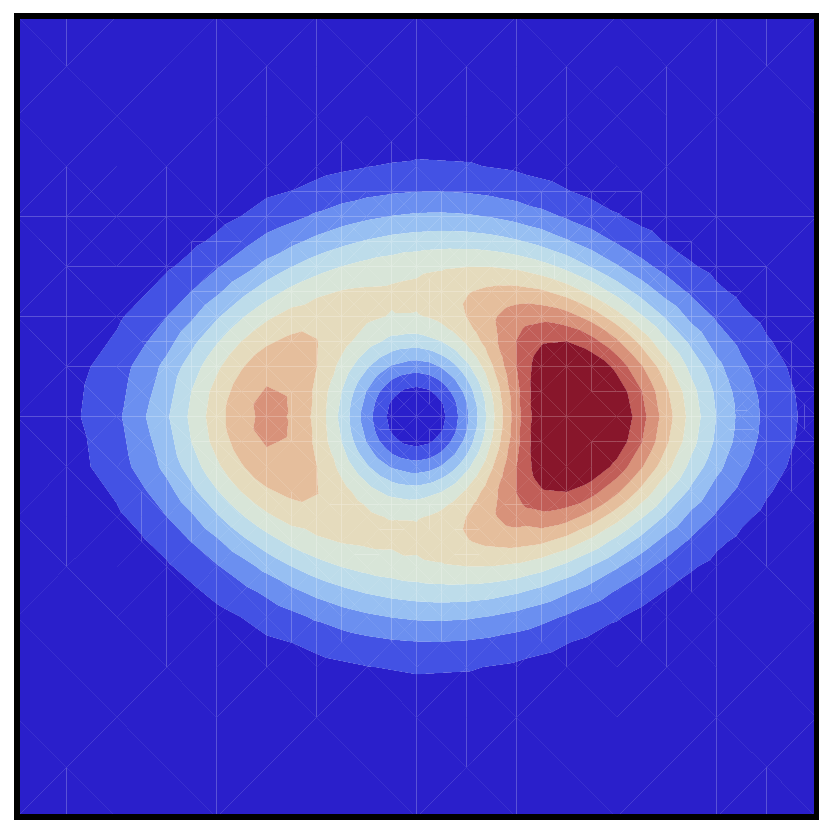}}
    \subfigure[$\lambda=10$]{\includegraphics[scale=0.28]{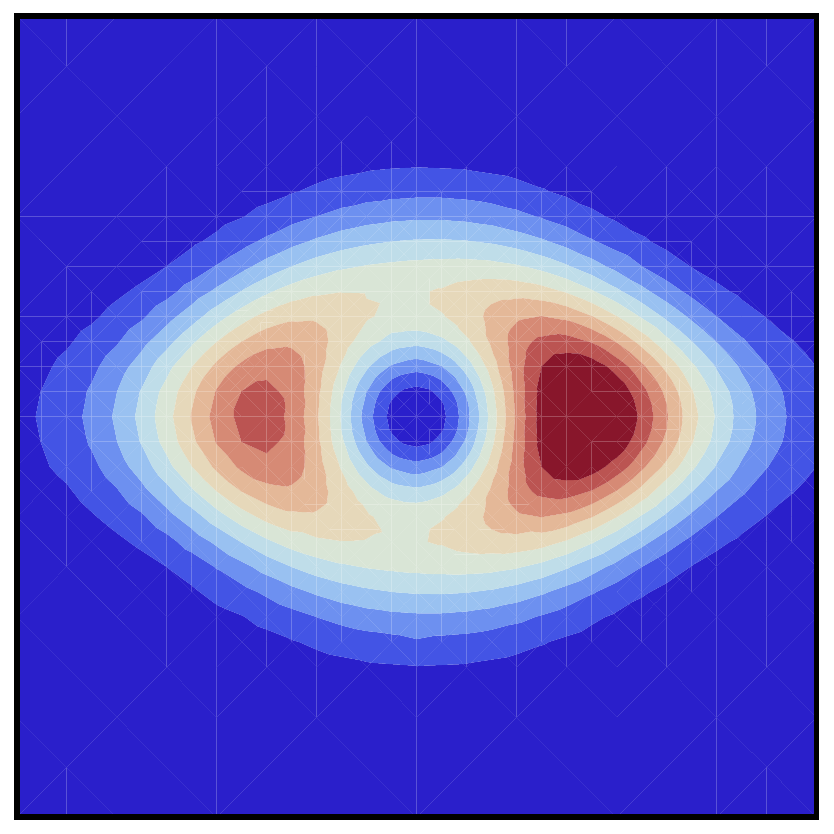}}
\end{center}
\caption{The $Q$-function corresponding to the solution of Eq.~(\ref{eq:DSe2}) with initial conditions $\theta_\varphi(0) = 0$, $\theta^\prime_\varphi(0) = e^{-i \varphi}$.}\label{fig:f3}
\end{figure*}

In the very special case of $\lambda=1$ the solution of the equation (\ref{eq:DSe}) reads as (see \cite{ps-55-528})
\begin{equation}\label{eq:ncs}
    |\psi\rangle = \mathcal{N}\left(|0\rangle + \sum^{+\infty}_{n=1} \frac{1}{\sqrt{n!}}
    \frac{\gamma^n}{g(0) \ldots g(n-1)} |n\rangle\right),
\end{equation}
provided that all the numbers $g(n)$, $n = 0, 1, \ldots$ are not equal to zero. As an example let us take $g(\hat{n}) = \hat{n}$. In this case Eq.~(\ref{eq:DSe}) reads as
\begin{equation}\label{eq:DSe2}
    (1+\lambda) r e^{i \varphi} \frac{d^2 \psi_\varphi(r)}{dr^2} +
    (1-\lambda) r^2 e^{-i \varphi} \frac{d \psi_\varphi(r)}{dr} =
    \beta \psi_\varphi(r).
\end{equation}
The constraints (\ref{eq:nic}) on the initial conditions in this case simply read as $\beta\psi_\varphi(0) = 0$, thereby if $\beta \not= 0$ then $\psi_\varphi(0) = 0$ and if $\beta = 0$ then both the derivatives $\psi_\varphi(0)$ and $\psi^\prime_\varphi(0)$ can be arbitrary. Since it is a second-order equation and if one of its initial conditions is fixed, then its (normalized) solution is unique. The $Q$-function of the solution is shown in Fig.~\ref{fig:f3}. In the case of $\lambda=1$ it is possible to solve Eq.~(\ref{eq:DSe2}) analytically, but we cannot directly use the expression given in Eq.~(\ref{eq:ncs}) since in this case $g(0)=0$. If $\beta \not=0 $ then the solution for $\lambda=1$ reads as follows:
\begin{equation}
    |\psi\rangle = {}_0F_2(;1, 2; |\beta|^2/4)^{-1/2} \  {}_0\tilde{F}_1(;2;\beta\hat{a}^\dagger/2) |1\rangle,
\end{equation}
where ${}_0\tilde{F}_1(;b;z)$ is the regularized confluent hypergeometric function defined via
\begin{equation}
    {}_0\tilde{F}_1(;b;z) = \frac{{}_0F_1(;b;z)}{\Gamma(b)}, \quad 
    {}_0F_1(;b;z) = \sum^{+\infty}_{k=0}\frac{1}{(b)_k} \frac{z^k}{k!}.
\end{equation}
If $\beta=0$ then the solution is just a linear combination of the first two Fock states $|\psi\rangle = c_0|0\rangle + c_1|1\rangle$, hence the solution is not unique.

\section{Summary and Conclusions}

In the present paper we have studied generalized squeezed states, which are based on the minimum of the uncertainty relation for two general Hermitian operators. The problem of finding the related generalized squeezed states has been reduced to solving ordinary differential equations which are obtained  in the Fock-Bargmann representation. These equations can be solved analytically only in some special cases, such as quadrature squeezing and amplitude-squared squeezing. For more general cases we have developed techniques to solve the differential equations numerically.

To illustrate the power of our approach, we have studied two examples of generalized squeezed states 
of types, which to our best knowledge have not been considered so far. The first type is a nonlinear generalization of the quadrature squeezed states, where in the definition of the quadrature operators
the annihilation operator is replaced by a function of the latter. The second type is a deformed nonlinear squeezed state, which is defined on the basis of the quadratures of a deformed algebra. 
For both types of generalized squeezed states we illustrate their properties by calculating the phase-space distributions, i.e. the $Q$-functions. 

In conclusion, we have studied  several types of generalized squeezed states.
The reduction of the quantum noise level of such states is related to different types of Hermitian operators. In this sense such kinds of squeezed states may become interesting when one wants to improve special measurement schemes. In such cases it may be of interest to find the best squeezed states in relation to the observable to be detected by a given device. When knowing the measurement scheme, our method is useful to characterize the squeezed states which are adjusted to the observation scheme. This may be a first step towards the preparation of these states and their applications for optimized measurements at a reduced level of quantum-noise. 

\appendix

\section{Normal ordering}

\label{app:NOa}

In this appendix we prove the following relation:
\begin{equation}\label{eq:no}
    g(\hat{a}) f(\hat{a}^\dagger) = \sum^{+\infty}_{k=0} \frac{1}{k!} f^{(k)}(\hat{a}^\dagger) g^{(k)}(\hat{a}),
\end{equation}
where $f(z)$ and $g(z)$ are entire functions. From the ubiquitous bosonic commutation relation $[\hat{a}, \hat{a}^\dagger] = 1$ it is easy to get the equality
\begin{equation}\label{eq:af}
    \hat{a}^n f(\hat{a}^\dagger) = \sum^n_{k=0} \binom{n}{k} f^{(k)}(\hat{a}^\dagger)
    \hat{a}^{n-k},
\end{equation}
by induction. Then we can calculate the left-hand side of Eq.~(\ref{eq:no}), assuming $g(x) = \sum^{+\infty}_{n=0} g_n x^n$
\begin{equation}
\begin{split}
    g(\hat{a}) f(\hat{a}^\dagger) &=
    \sum^{+\infty}_{n=0} \sum^n_{k=0} \binom{n}{k} f^{(k)}(\hat{a}^\dagger) g_n
    \hat{a}^{n-k} \\
    &= \sum^{+\infty}_{k=0} \frac{1}{k!} f^{(k)}(\hat{a}^\dagger)
    \sum^{+\infty}_{n=k} \frac{n!}{(n-k)!} g_n \hat{a}^{n-k} \\
    &= \sum^{+\infty}_{k=0} \frac{1}{k!} f^{(k)}(\hat{a}^\dagger) g^{(k)}(\hat{a}), 
\end{split}
\end{equation}
which gives us the desired result. Here we used the following symbolical rule for double sums:
\begin{equation}
    \sum^{+\infty}_{n=0} \sum^n_{k=0} = \sum^{+\infty}_{k=0} \sum^{+\infty}_{n=k}.
\end{equation}
If the function $f(z)$ is real then the commutator $[f(\hat{a}), f(\hat{a}^\dagger)]$ reads as
\begin{equation}
    [f(\hat{a}), f(\hat{a}^\dagger)] = \sum^{+\infty}_{k=1} \frac{1}{k!} f^{(k)}(\hat{a})^\dagger f^{(k)}(\hat{a}),
\end{equation}
so that it is nonnegative.

\section{Amplitude squared squeezing}

\label{app:ASS}

In this appendix we prove that the states given in the Fock-Bargmann representation  (\ref{eq:ASs}) can be rewritten in the form of Eq.~(\ref{eq:ASp}). First of all, we show that the functions in Eq.~(\ref{eq:ASs}) do not depend on the choice of the sign for the parameter $c$ in Eq.~(\ref{eq:bc}). There are two possible values for $c$, which differ in sign, denoted here as $c_+$ and $c_-$ so that $c_+ = -c_-$. To $c_+$ and $c_-$ correspond two values for $b$, $b_+$ and $b_-$ respectively. As one can easily see from Eq.~(\ref{eq:bc}) they are related as $b_+ = 1/2 - b_-$. It is well known that the Kummer function ${}_1F_1$ satisfies to the following relation:
\begin{equation}
    {}_1F_1(b; c; -z) = e^{-z} {}_1F_1(c-b; c; z).
\end{equation}
Using this relation we can write
\begin{equation}
\begin{split}
    \psi_{\mathrm{e}}(b_-; c_-; z) &= e^{c_+ z^2} {}_1F_1\left(b_-; \frac{1}{2}; -2c_+z^2\right) \\
    &= e^{-c_+ z^2} {}_1F_1\left(\frac{1}{2}-b_-; \frac{1}{2}; 2c_+z^2\right) \\
    &= \psi_{\mathrm{e}}(b_+; c_+; z).
\end{split}
\end{equation}
The same is also true for the odd function $\psi_{\mathrm{o}}(b; c; z)$.

Now we prove the following relation:
\begin{equation}\label{eq:psieven}
\begin{split}
    S(\xi) &{}_1F_1\left(b; \frac{1}{2}; v\hat{a}^{\dagger 2}\right) |0\rangle =\frac{1}{\sqrt{\mu}(1-2\zeta^*v)^b} \\
    &\times e^{-\zeta \hat{a}^{\dagger 2}/2}
    {}_1F_1\left(b; \frac{1}{2}; \frac{v\hat{a}^{\dagger 2}}{\mu(\mu-2\nu^*v)}\right) |0\rangle,
\end{split}
\end{equation}
where the parameters $\mu$, $\nu$ and $\zeta$ read as
\begin{equation}
    \quad \mu = \cosh |\xi|, \quad \nu = e^{i \arg\xi} \sinh|\xi|, \quad \zeta = \frac{\nu}{\mu}.
\end{equation}
Note that the squeezing operator can be written as $S(\xi) = e^{\xi^* \hat{K}_--\xi\hat{K}_+}$, where the operators $\hat{K}_\pm$ and $\hat{K}_0$ are defined to be
\begin{equation}
    \hat{K}_+ = \frac{\hat{a}^{\dagger 2}}{2}, \quad
    \hat{K}_- = \frac{\hat{a}^2}{2}, \quad
    \hat{K}_0 = \frac{\hat{n}+1/2}{2}.
\end{equation}
These operators satisfy to the following commutation relations:
\begin{equation}
    [\hat{K}_0, \hat{K}_\pm] = \pm \hat{K}_\pm, \quad
    [\hat{K}_-, \hat{K}_+] = 2\hat{K}_0,
\end{equation}
and these three operators generate the Lie algebra of the group $\mathrm{SU}(1, 1)$. The normal form of the squeezing operator is given by (cf. \cite{perelomov})
\begin{equation}
\begin{split}
    S(\xi) &= e^{-\zeta\hat{K}_+} \left(\frac{1}{\mu}\right)^{2\hat{K}_0} e^{\zeta^*\hat{K}_-} \\
    &= \frac{1}{\sqrt{\mu}} : 
    \exp\left(-\frac{\zeta\hat{a}^{\dagger 2}}{2} + \frac{\zeta^*\hat{a}^2}{2}+\frac{\mu-1}{\mu}\hat{n}\right):.
\end{split}
\end{equation}
We will transform the left-hand side of (\ref{eq:psieven}) step by step. First, we must calculate the expression $|\tilde{\psi}\rangle = e^{\zeta^* \hat{a}^2/2} {}_1F_1\left(b; 1/2; v\hat{a}^{\dagger 2}\right) |0\rangle$ and this is the only nontrivial step in the whole process. According to the relation (\ref{eq:no}) we can write it as
\begin{equation}\label{eq:d1}
    |\tilde{\psi}\rangle = \sum^{+\infty}_{n=0} \frac{1}{n!} {}^{}_1F_1^{(n)}\left(b; \frac{1}{2}; v \hat{a}^{\dagger 2}\right) \left.\frac{d^n e^{\zeta^* z^2/2}}{dz^n}\right|_{z=0}|0\rangle.
\end{equation}
The derivative of the exponent at zero reads as follows:
\begin{equation}
    \left.\frac{d^n e^{\zeta^* z^2/2}}{dz^n}\right|_{z=0} =
    \begin{cases}
        0 & \mathrm{if}\ n\ \mathrm{is\ odd} \\
        (\zeta^*/2)^{n/2} \frac{n!}{(n/2)!} & \mathrm{if}\ n\ \mathrm{is\ even},
    \end{cases}
\end{equation}
thereby only terms with even $n$ survive in the sum (\ref{eq:d1}). An even order derivative of the Kummer function reads as
\begin{equation}
    {}_1F^{(2k)}_1\left(b; \frac{1}{2}; v z^2\right) =
    (4v)^k (b)_k\ {}_1F_1\left(b+k; \frac{1}{2}; v z^2\right).
\end{equation}
Now we can further proceed with the expression for $|\tilde{\psi}\rangle$
\begin{equation}
    |\tilde{\psi}\rangle = \sum^{+\infty}_{k=0} (b)_k \frac{(2 \zeta^* v)^k}{k!}
    {}_1F_1\left(b+k; \frac{1}{2}; v \hat{a}^{\dagger 2}\right) |0\rangle.
\end{equation}
According to the definition (\ref{eq:kummer}) of the Kummer function we have the equality
\begin{equation}
    {}_1F_1\left(b+k; \frac{1}{2}; w\right) = \sum^{+\infty}_{m=0}
    \frac{(b+k)_m}{(1/2)_m} \frac{w^m}{m!},
\end{equation}
and upon substituting it into the previous relation we finally get
\begin{equation}
\begin{split}
    |\tilde{\psi}\rangle &=
    \sum^{+\infty}_{k, m=0} \frac{(2\zeta^*v)^k}{k!} \frac{(b)_{k+m}}{(1/2)_m} \frac{(v\hat{a}^{\dagger 2})^m}{m!}|0\rangle \\
    &= \frac{1}{(1-2\zeta^*v)^b} \sum^{+\infty}_{m=0} \left(\frac{v\hat{a}^{\dagger 2}}{1-2\zeta^*v}\right)^m \frac{(b)_m}{(1/2)_m m!} |0\rangle \\
    &= \frac{1}{(1-2\zeta^*v)^b} {}_1F_1\left(b; \frac{1}{2}; \frac{v\hat{a}^{\dagger 2}}{1-2\zeta^*v}\right)|0\rangle. 
\end{split}
\end{equation}
Here we used the relation $(b)_k (b+k)_m = (b)_{k+m}$ and the following equality:
\begin{equation}
    \sum^{+\infty}_{k=0} (b)_{k+m} \frac{x^k}{k!} = \frac{(b)_m}{(1-x)^{b+m}}.
\end{equation}
We have just obtained the following expression for $|\tilde{\psi}\rangle$:
\begin{equation}
    |\tilde{\psi}\rangle = \frac{1}{(1-2\zeta^*v)^b} {}_1F_1\left(b; \frac{1}{2}; \frac{v}{1-2\zeta^*v} \hat{a}^{\dagger 2}\right) |0\rangle.
\end{equation}
Then we must apply the operator $(1/\mu)^{\hat{n}+1/2}$ to $|\tilde{\psi}\rangle$. As we already mentioned, this is equivalent to scaling the argument of the corresponding function in the Fock-Bargmann representation. Applying this scaling to the left-hand side of the previous equation and multiplying by $e^{-\zeta\hat{a}^{\dagger 2}/2}$, we finally arrive to Eq.~(\ref{eq:psieven}).

Now we must find $\xi$ and $v$ such that the right-hand side of Eq.~(\ref{eq:psieven}) is exactly the even state defined in Eq.~(\ref{eq:ASp}). It is easy to see that $\xi$ and $v$ defined in Eq.~(\ref{eq:ASxv}) have this property. This finishes the proof of the representation of even and odd states in the Fock-Bargmann representation given by Eq.~(\ref{eq:ASs}) in the form of Eq.~(\ref{eq:ASp}).

Now we calculate the normalization of the even and odd states under study. It is much easier to do when these states are represented in the form of Eq.~(\ref{eq:ASp}), for this reason we have transformed the states into this form. The normalization $\mathcal{N}_{\mathrm{e}}$ of the even state can be calculated as
\begin{equation}
    \mathcal{N}^{-2}_{\mathrm{e}} = \sum^{+\infty}_{n=0} \frac{1}{n!} {}_1F_1\left(b; \frac{1}{2}; v z^2\right)^{(n)}_{z=0}
    {}_1F_1\left(b; \frac{1}{2}; v z^2\right)^{(n)}_{z=0}.
\end{equation}
The derivatives in this expression read as
\begin{equation}
    {}_1F_1\left(b; \frac{1}{2}; v z^2\right)^{(n)}_{z=0} =
    \begin{cases}
        0 & \mathrm{if}\ n\ \mathrm{is\ odd} \\
        (4v)^{n/2} (b)_{n/2} & \mathrm{if}\ n\ \mathrm{is\ even},
    \end{cases}
\end{equation}
thereby the normalization $\mathcal{N}_{\mathrm{e}}$ can be written as follows:
\begin{equation}
    \mathcal{N}^{-2}_{\mathrm{e}} = \sum^{+\infty}_{m=0} (b)_m (b^*)_m \frac{(16|v|^2)^m}{(2m)!}.
\end{equation}
Using the relation $(2m)! = 4^m m! \left(1/2\right)_m$, we finally get Eq.~(\ref{eq:ASn}). In the same way one can obtain the normalization $\mathcal{N}_{\mathrm{o}}$ of the odd state.

\section{Normal ordering of photon-number operator functions}

\label{app:NO}

In this appendix we prove the following equality:
\begin{equation}\label{eq:Af}
    g(\hat{n}) = :(e^{\hat{n} \Delta}g)(0):,
\end{equation}
where $f(z)$ is an entire function. Let us start with the expression for normally-ordered powers of the photon number operator
\begin{equation}
    :\hat{n}^m: = \hat{n} (\hat{n}-1) \ldots (\hat{n}-m+1) = \sum^m_{k=0} s(m, k) \hat{n}^k,
\end{equation}
where $s(m, k)$ are the signed Stirling numbers of the first kind. One can invert this relation and express ordinary powers of $\hat{n}$ in terms of normally-ordered ones
\begin{equation}
    \hat{n}^m = \sum^m_{k=0} S(m, k) :\hat{n}^k:,
\end{equation}
where $S(m, k)$ are the Stirling numbers of the second kind. Explicitly these numbers read as
\begin{equation}\label{eq:s2}
    S(m, k) = \frac{1}{k!} \sum^k_{i=0} (-1)^{k-i} \binom{k}{i} i^m.
\end{equation}
This is the number of ways to partition a set of $m$ elements into $k$ non-empty subsets and hence it is equal to zero if $m<k$ (see \cite{abramowitz}). Using the Taylor expansion $g(z) = \sum^{+\infty}_{m=0} g_m z^m$ of the function $f(z)$ and replacing $z$ with $\hat{n}$ we can write the operator $g(\hat{n})$ as
\begin{equation}
\begin{split}
    g(\hat{n}) &= \sum^{+\infty}_{m=0} g_m \hat{n}^m =
    \sum^{+\infty}_{m=0} g_m \sum^m_{k=0} S(m, k) :\hat{n}^k: \\
    &= \sum^{+\infty}_{k=0} \left(\sum^{+\infty}_{m=k} S(m, k) g_m \right) :\hat{n}^k: =
    \sum^{+\infty}_{k=0} F_k :\hat{n}^k:.
\end{split}
\end{equation}
Now we have to calculate $F_k$ defined via
\begin{equation}
    F_k = \sum^{+\infty}_{m=k} S(m, k) g_m = \frac{1}{k!} \sum^k_{i=0} (-1)^{k-i} \binom{k}{i} \sum^{+\infty}_{m=k} g_m i^m.
\end{equation}
The second sum can be represented as follows:
\begin{equation}
    \sum^{+\infty}_{m=k} g_m i^m = g(i) - \sum^{k-1}_{m=0} g_m i^m,
\end{equation}
so that $F_k$ can now be written as
\begin{equation}
    F_k = \frac{1}{k!} \sum^k_{i=0} (-1)^{k-i} \binom{k}{i} g(i) - \sum^{k-1}_{m=0} S(m, k) g_m.
\end{equation}
It has been mentioned above that $S(m, k)=0$ for $m<k$, so the last sum in this expression is zero, and finally we have
\begin{equation}
    F_k = \frac{1}{k!} \sum^k_{i=0} (-1)^{k-i} \binom{k}{i} g(i) =
    \frac{1}{k!} (\Delta^k g)(0),
\end{equation}
which completes the proof of Eq.~(\ref{eq:Af}).

\section{The relations for the initial conditions}

\label{app:IC}

In this appendix we prove the relation (\ref{eq:nic}) for the initial conditions for Eq.~(\ref{eq:DSd}). It is easy to see that the derivatives at zero of the functions $C_k(r)$ defined in Eq.~(\ref{eq:DSC}) read as
\begin{equation}\label{eq:C}
    C^{(m)}_k(0) =
    \begin{cases}
        (1+\lambda)(\Delta^{k-1}g)(0) e^{i \varphi} & m = k-1 \\
        (k+1)(1-\lambda)(\Delta^k g)(0) e^{-i \varphi} & m = k+1 \\
        0 & m \not= k \pm 1
    \end{cases}
\end{equation}
Differentiating Eq.~(\ref{eq:DSd}) $m$ times at $r=0$ we get the following relation:
\begin{equation}
    \sum^{+\infty}_{k=0} \sum^m_{j=0} \left(\begin{array}{c}m\\j\end{array}\right) C^{(j)}_k(0) \psi^{(k+m-j)}_\varphi(0)
    = \beta \psi^{(m)}_\varphi(0).
\end{equation}
Using the expressions (\ref{eq:C}) this relation can be simplified as
\begin{equation}
\begin{split}
    &\sum^m_{j=0} \left(\begin{array}{c}m\\j\end{array}\right) \bigl(C^{(j)}_{j+1}(0) \psi^{(m+1)}_\varphi(0) +
    C^{(j)}_{j-1}(0) \psi^{(m-1)}_\varphi(0)\bigr) \\
    &= \beta \psi^{(m)}_\varphi(0).
\end{split}
\end{equation}
The coefficient in front of the derivative $\psi^{(m+1)}_\varphi(0)$ can be further simplified as follows:
\begin{equation}
\begin{split}
    \sum^m_{j=0} \left(\begin{array}{c}m\\j\end{array}\right) &C^{(j)}_{j+1}(0) = (1+\lambda) e^{i \varphi}
    \sum^m_{j=0} \left(\begin{array}{c}m\\j\end{array}\right) (\Delta^j g)(0) \\
    &=(1+\lambda) e^{i \varphi} ((1+\Delta)^m g)(0) \\
    &= (1+\lambda) e^{i \varphi} (E^m g)(0) = (1+\lambda) g(m) e^{i \varphi}, 
\end{split}
\end{equation}
where $E = 1+\Delta$ is the step operator which acts as $(Eg)(n) = g(n+1)$. In the same way one can calculate the coefficient in front of $\psi^{(m-1)}_\varphi(0)$ and get
\begin{equation}
    \sum^m_{j=0} \left(\begin{array}{c}m\\j\end{array}\right) C^{(j)}_{j-1}(0) = m(1-\lambda) g(m-1) e^{-i\varphi}.
\end{equation}
This completes the proof of the relation (\ref{eq:nic}).

\end{document}